\documentclass[pdflatex,sn-mathphys-num]{sn-jnl}

\usepackage{graphicx}%
\usepackage{multirow}%
\usepackage{amsmath,amssymb,amsfonts}%
\usepackage{amsthm}%
\usepackage{mathrsfs}%
\usepackage[title]{appendix}%
\usepackage{xcolor}%
\usepackage{textcomp}%
\usepackage{manyfoot}%
\usepackage{booktabs}%
\usepackage{algorithm}%
\usepackage{algorithmicx}%
\usepackage{algpseudocode}%
\usepackage{listings}%
\usepackage[draft]{changes}

\usepackage{todonotes}
\usepackage{comment}
\usepackage{color, colortbl}	

\usepackage{adjustbox}
\NewDocumentCommand{\rot}{O{90} O{1em} m}{\makebox[#2][l]{\rotatebox{ #1}{#3}}}%

\definecolor{mygray}{gray}{0.9}
\newcolumntype{g}{>{\columncolor{mygray}}c}

\usepackage{changes}
\usepackage{xcolor}
\usepackage{csquotes}
\definechangesauthor[name=Jörg, color=magenta]{Jhe}
\definechangesauthor[name=Bettina, color=brown]{BJ}
\definechangesauthor[name=All, color=orange]{ALL}

\usepackage{ulem}

\theoremstyle{thmstyleone}%
\theoremstyle{thmstyletwo}%

\theoremstyle{thmstylethree}%

\begin{document}

\title[Quantum MOOCs at Open HPI]{
Introducing Quantum Information and Computation to a Broader Audience with MOOCs at OpenHPI }


\author[1]{\fnm{Gerhard} \sur{Hellstern}}\email{gerhard.hellstern@dhbw-stuttgart.de}

\author[2]{\fnm{Jörg} \sur{Hettel}}\email{joerg.hettel@hs-kl.de}

\author*[3]{\fnm{Bettina} \sur{Just}}\email{bettina.just@mni.thm.de}
\equalcont{All authors contributed equally to this work.}

\affil[1]{\orgdiv{Center of Finance}, \orgname{Cooperative State University of Baden-Württemberg (DHBW)}, \orgaddress{\city{Stuttgart}, \country{Germany}}}

\affil[2]{\orgdiv{Department of Computer Sciences and Microsystems Technology}, \orgname{University of Applied Sciences Kaiserslautern}, \orgaddress{\country{Germany}}}

\affil[3]{\orgdiv{FB MNI}, \orgname{Technische Hochschule Mittelhessen (THM)}, \city{Gießen}, \country{Germany}}

\abstract{
Quantum computing is an exciting field with high disruptive potential, but very difficult to access. For this reason,
many approaches to teaching quantum computing are being developed worldwide.
  This always raises questions about the didactic concept, the content actually taught, and how to measure the success of the teaching concept.
In 2022 and 2023, the authors   taught a total of nine two-week MOOCs (massive open online courses) with different possible learning paths on the Hasso Plattner Institute's OpenHPI platform. The   purpose of the platform is to make computer science education available to everyone free of charge. The nine quantum courses form a self-contained curriculum. A total of more than 17{,}000   course attendances have been taken by about 7400   natural persons, and the number is still rising.
This paper presents the course concept and evaluates the anonymized
data on the background of the participants, their behaviour in the courses, and their learning success.
 This paper is the first to analyze
such a large dataset of MOOC-based quantum computing education.
The summarized results are a heterogeneous personal background of the participants biased towards IT professionals, a majority following the didactic recommendations, and a high success rate, which is strongly correlatated   with following the didactic recommendations. The amount of data from such a large group of quantum computing learners  provides many avenues for further research in the field of quantum computing education.
The analyses show that the MOOCs are a low-threshold concept for getting into quantum computing. It was very well received by the participants. The concept can serve as an entry point and guide for the design of quantum computing courses.}




\maketitle

\section{Introduction}\label{introduction}

Quantum technologies are considered to be one of the most important technologies
of the near future. Recently, the sub-area of quantum computing and quantum
communication has increasingly come into focus, see Sect.~\ref{literature_review} for a brief glance at the literature. Many companies
have launched initial innovation projects in order not to miss out on these
technologies. However, a major problem is the lack of skilled workers,
particularly in the field of quantum computing. Universities have now set
up relevant courses, but the number of graduates will not meet the expected
demand for specialists. Politicians have recognized this and have launched
various programs. The Federal Ministry of Education and Research in Germany
(BMBF) has therefore initiated the  ``Quantum Future
of Education'' funding measure for 2021. The aim was to  ``develop
innovative, interdisciplinary concepts and programs for education and training
in quantum technologies''.

As part of this project, an innovative MOOC (Massive Open Online Courses)
curriculum on topics related to quantum computing was developed by the
authors of the present paper in cooperation with the Hasso Plattner Institute
(HPI). The curriculum consists of nine courses and is explicitly addressed
  to an audience with little prior knowledge of the
subject. The aim was to pick up participants at their level of knowledge
and use innovative didactic concepts to convey a sound understanding of
various areas of quantum computing. The focus was not on quantum physics
or mathematics but on applications of the underlying concepts. As this
was a BMBF project and was primarily aimed at a German audience, the MOOC
content is in German with an ongoing process of English subtitles being
added. The knowledge acquired in the MOOCs is intended to serve as a starting
point for further individual specialized professional training.

To make it easier for participants to get started, they are offered various
learning paths. This allows participants to put together an individual
learning profile depending on their previous knowledge and interests.
On January 17th, 2024, when the data for this paper was exported,
  7413 participants had attended
the MOOCs,
who had taken an overall of 17{,}157 courses in the curriculum. The
topic of this paper is to analyze the learning behavior of these participants
within the MOOCs. Therefore, the following research questions are addressed:
\begin{enumerate}[RQ 2:]
\item[RQ 1:] What kind of audience attended the MOOCs? What are
the specific backgrounds and interests of the participants?
\item[RQ 2:]
 What specific topics or subject areas did the participants choose when selecting the courses offered?
Which topics are the participants particularly interested in?
\item[RQ 3:] Which competences could be acquired by the participants
through the course concept? To what extent did the course concept support
the participants?
\end{enumerate}

This paper is organized as follows. Section~\ref{literature_review} provides
a brief (and necessarily selective) overview of the enormous amount of
literature on quantum computing education. Section~\ref{curriculum} presents
the didactic concept of the curriculum under consideration in this paper,
as well as the course's contents. The content is also categorized with
respect to the European Competence Framework for Quantum Technologies
\cite{franziska_greinert_towards_2023}. Section~\ref{results} evaluates
the data collected by HPI during the courses with respect to the three
research questions mentioned above. In Sect.~\ref{conclusion}, based
on the results of Sect.~\ref{results}, we draw conclusions
  concerning the research questions, the
lessons learned, and point out topics for further research.

\section{Literature review}
\label{literature_review}

While quantum technologies have recently gained prominence in educational
research, the investigation into quantum physics pedagogy is longstanding
\cite{philipp_bitzenbauer_quantum_2021}. Bitzenbauer's literature review
spanning 2000 to 2021 encapsulates 1520 works on this subject.

Research in imparting skills and competencies for quantum technologies
has primarily centered on university education, particularly Bachelor's
and Master's degree programs \cite{clarice_d_aiello_achieving_2021}. Aiello's
work discusses 18 programs dedicated to training in quantum information
science and engineering, emphasizing the necessity for corresponding investments
across various stakeholders.

A targeted approach to teaching quantum technology and computing at US
universities, especially within Historically Black Colleges and Universities
(HBCUs), have been noted \cite{kayla_lee_ibm-hbcu_2021}. Lee's research
addresses issues of Black representation and workforce diversity in quantum
information science and engineering.

Bungum's study focuses on developing a quantum course for master's level
information technology students, drawing insights from participant interviews
\cite{bungum_what_2022}. Similarly, Stump's analysis explores student difficulties
and misunderstandings regarding quantum topics
\cite{emily_m_stump_context_2023}.

An evaluation of interdisciplinary approaches to teaching quantum information
science is detailed in Meyer's work, highlighting the influence of instructors
and the need for diverse perspectives
\cite{josephine_c_meyer_todays_2022}. The presentation of a 1D quantum
simulation and visualization tool by Zaman Ahmed aims to enhance understanding
of quantum phenomena from the students' perspective
\cite{shaeema_zaman_ahmed_student_2022}.

Delgado's examination of the effects of the COVID-19 pandemic on quantum
information science education underscores the evolving landscape of online
teaching modalities \cite{francisco_delgado_extending_2023}. Lastly, Hasanovic's
account of the NSF-funded EdQuantum project highlights efforts to develop
a curriculum for future quantum technicians
\cite{moamer_hasanovic_quantum_2023}.

In contrast to tertiary education, educational research on quantum topics
  focussing on elementary and secondary schools
just started a few years ago.

The study of Bondani et al. \cite{bondani_introducing_2022} is initiated
by the European QTEdu CSA project, a component of the  ``Quantum
Flagship'', wherein courses on quantum physics and the application of quantum
technologies are devised, implemented, and evaluated among approximately
250 students. Freericks et al. \cite{Freericks_2017} already 2017 developed
MOOCs for teaching quantum mechanics to over 28.000 non-scientists.

The work by Angara et al. \cite{Angara2020}, Santanassi et al.
\cite{Satanassi2022,Satanassi2023}, Pospiech \cite{Pospiech2021}, Sutrini
et al. \cite{Sutrini2022} investigate the possibilities to include basic
quantum algorithms (e.g. teleportation) into physics classes. In addition,
Sutirini et al. \cite{Sutrini2022,Sutrini2023a} investigated how training
courses for physics teachers can be set up.

The training   in both school and university settings
on quantum technologies ultimately aims to address the anticipated global
demand for skilled professionals in these domains. This foreseen need is
compellingly expounded upon by Venegas-Gomez and Plunkett
\cite{araceli_venegasgomez_quantum_2020,thomas_plunkett_survey_2020}.

Several works specifically tackle the imperative of cultivating a requisite
national and international workforce in quantum domains. Fox et al.
\cite{michael_f_j_fox_preparing_2020} conclude that the commercialization
of quantum technologies necessitates an adequately trained workforce, drawing
insights from a qualitative study involving 21 US companies. Asfaw et al.
\cite{abraham_asfaw_building_2022} chart a roadmap for developing a commensurate
workforce through the establishment of a quantum engineering education
program, informed by a survey of 480 researchers across US universities,
government agencies, industry, and research laboratories.

Although the aforementioned studies primarily center on the US landscape
works such as those by Greinert (nee Gerke) et al.
\cite{f_gerke_requirements_2022,franziska_greinert_future_2023,franziska_greinert_towards_2023}, contribute to the formulation and presentation of a European competence
framework. This framework underpins the development of training programs
aimed at nurturing the future quantum workforce. Such initiatives were
integral to the European project QTEdu CSA, which began to establish a
framework for second-generation quantum technologies - activities that
are now continuing as part of QUCATS, the current Quantum Flagship CSA.

The education and training of specialists in quantum technologies necessitate
broader dissemination of these subjects in academic institutions alone.
Engagement with individuals already employed in companies is imperative.
Additionally, initiatives must cater to students and schoolchildren lacking
adequate educational provisions thus far.

Quantum-related online courses were introduced at the European Research
Center CERN in 2020, as evaluated in Combarro's report
\cite{elias_f_combarro_report_2021}, which examines participant demographics
and course outreach.

Maldonado-Romo's study reports on a Spanish-language quantum online event
featuring introductory workshops and hackathons, engaging 220 participants
from Latin America, with two-thirds being beginners
\cite{alberto_maldonado-romo_quantum_2022}.

The integrated approach presented in the present paper, which encompasses
fundamental quantum concepts, quantum computing, and quantum cryptography,
is echoed in Aithal's work \cite{p_s_aithal_advances_2023}. Notably, Aithal
underscores synergies among various topics within the realm of Information,
Communication, and Computing Technologies (ICCT).

\section{The OpenHPI Quantum Channel and the curriculum on Quantum Computing}\label{curriculum}

Since 2012, the Hasso Plattner Institute (HPI)  has operated the online educational platform OpenHPI (\url{https://open.hpi.de}).
The platform offers free access to MOOCs in the field of information technology.
A MOOC at OpenHPI usually consists   of around
20 concise videos (approximately 10-15 minutes each), 10 self-tests to
assess participant learning progress and one final examination test. The
self-tests and the final examination tests are carried out online and are
multiple-choice tests with more than one choice possible for the questions.
The self-tests consist of about 10 questions each, the final exam of 20
questions.

Each course is conducted within a specified timeframe. Participants can
repeat self-tests as needed, but the final exam can only be taken once
online at the end of the time frame and has to be finished within 120 minutes.

The time frame of a MOOC normally is   two weeks
since studies in the literature support  the determination that
a two-week module with integrated self-tests and a final exam is optimal
\cite{Serth_2022}.

Throughout the course's timeframe, participants can engage in discussions
and pose questions within a forum moderated by the instructor. Answers
to questions and comments are usually provided within one day. A course
can be completed in two ways. If participants have accessed more than 50\%
of the items (videos, self-tests, additional material), they receive a
certificate of attendance. If participants have more than 50\% of the points
achieved in the final examination, they receive a qualification certificate.
It should be noted that it is not necessary to access any of the items
to obtain a qualification certificate. It is sufficient to pass the final
exam.
These are the commen rules for all the courses at openHPI.

Even after the MOOCs have ended, the courses are still available to new
interested parties. However, the forums are closed, and the final exam
cannot be taken anymore. The videos can now be viewed on the platform without
registering. However, registered participants can access the self-test
and old forum entries. Participants can still complete the course successfully
by accessing 50 \% or more of the items. Then they will also receive a
certificate of attendance.

In the present paper, a distinction between so-called course learners and
self-learners is made. Course learners attend the course during the timeframe
and are able to interact with the course instructor and fellow students.
Self-learners register for the course after the end  of the
timeframe. They can only attend the course in a non-interactive
way.

In 2022, a dedicated channel for quantum computing was established on OpenHPI
(\url{https://open.hpi.de/channels/quantum}). The channel features 15 MOOCs addressing
various aspects of quantum computing. The kernel is an introductory curriculum
of nine interconnected and thematically coordinated MOOCs developed by
the authors of the present paper from June 2022 to July 2023. Concurrently,
six additional quantum MOOCs were developed as specialized, independent
courses. These courses cover diverse topics such as quantum computing for
school pupils, a general introduction to Qiskit, or advanced topics like
simulating quantum systems (Quantum Computing for Natural Sciences). The
present paper discusses and evaluates the curriculum of the nine interconnected
quantum MOOCs forming the kernel of the channel.

\subsection{Didactic Concept}\label{didactic_design}
The didactic concept aims to give participants with no prior
knowledge of mathematics or physics a sound introduction to quantum information.
The broad field is structured by three pillars  ``introduction'',
 ``cryptography'' and  ``algorithms''.
However, it is not required to work through these pillars one by one. Instead,
a series of self-contained learning paths is offered, so that participants
can set their own priorities and not have to deal with topics outside their
area of interest. In the end, they are able to use the acquired knowledge
to face the quantum computing topics occurring in their proper professional
practice.

\begin{figure}[h]
\centering
\includegraphics[width=0.5\textwidth]{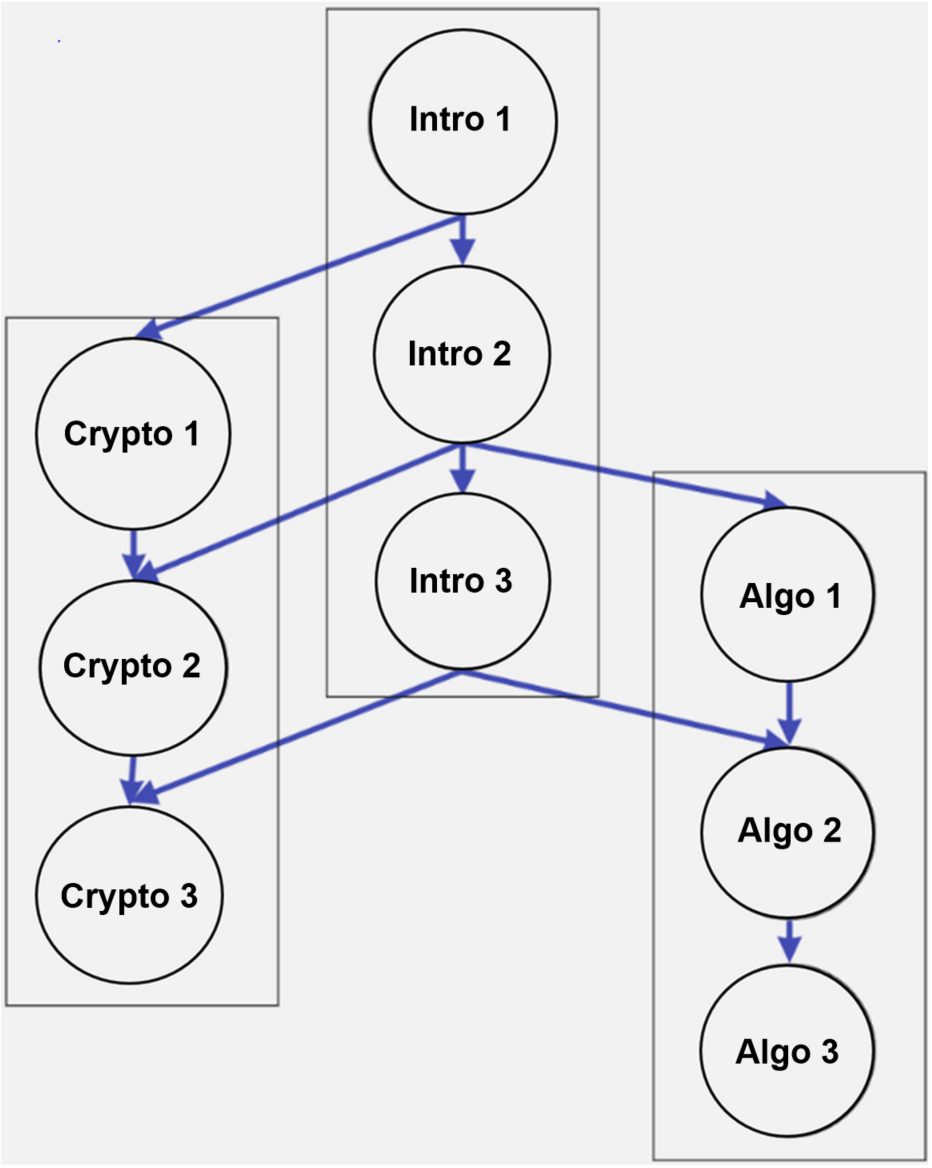}
\caption{Structure of the curriculum with nine courses and self-contained learning paths.}\label{QuantenkurseBauenAufeinanderAuf}
\end{figure}

The learning paths are illustrated in Fig.~\ref{QuantenkurseBauenAufeinanderAuf}. The nine courses are categorized
into three sections: three introductory courses (Intro 1, Intro 2, and
Intro 3), three courses focusing on quantum cryptography (Crypto 1, Crypto
2, and Crypto 3), and three courses on quantum algorithms (Algo 1, Algo
2, and Algo 3). Intro 1 does not require any prior knowledge. Intro 2 requires
Intro 1, but with Intro 1 alone it is already possible to understand Crypto
1. In general, it is possible to understand a course with the knowledge
of the courses prior to it in Fig.~\ref{QuantenkurseBauenAufeinanderAuf}. For instance, Algo 2 can be understood
with the knowledge of Intro 1, Intro 2, Intro 3 and Algo 1. Especially,
the pillars   about crypto and algorithms are distinct
one from the other, so participants who want to deepen their knowledge
in quantum cryptography need nothing to learn about (advanced) quantum
algorithms, and vice versa.
 A
proposed learning path is  a subset of the courses following the arrows in Fig.~\ref{QuantenkurseBauenAufeinanderAuf} and with each course
also containing   its predecessors. Information on
the structure of the curriculum and the structure of the learning paths
is given in the first video of each course, thus providing an orientation
on where the course is located in the curriculum, and also
  information on where to get prior knowledge if necessary for
a more advanced course in the curriculum. When useful, also the last video
of a course provides this information as a review and outlook. The modular
structure allows participants to customize their learning paths based on
their interests and pre-existing knowledge. For course-learners, the learning
path for all 9 courses was determined by the order of the courses' timeframes:
Intro 1, Crypto 1, Intro 2, Crypto 2, Algo 1, Intro 3, Crypto 3, Algo 2,
Algo 3. (Algo 2 and Algo 3 were merged together for practical reasons).
They could of course attend subsets of the courses, but in the timeframes'
order. Self-learners can work through the courses in any order of their
choice, but it is recommended to complete the previous courses in Fig.~\ref{QuantenkurseBauenAufeinanderAuf} before attending a new course.

In designing the individual courses, the authors drew on their experience of teaching at Universities of Applied Sciences. The aim here is always to assume as little prior knowledge as possible, minimize the theoretical background, use graphical representations, and activate the participants through integrated exercises. Repetitions of the same topic in different courses from different perspectives are also intentionally included to deepen the participants' understanding.\\

Special didactic features in the course pillars are as follows:
\begin{itemize}
\item In the introductory courses, the  ``quantum cube''
is used throughout to illustrate quantum register states and the effect
of quantum gates. This model was developed by Just
\cite{Just2020,Just2022} and since then has been studied and used as an
intuitive tool for visualizing quantum entanglement and quantum computation
\cite{seegerer2021quantum,Fraunhofer2022,bley2024visualizing}. It was
applied to the more advanced topics of quantum oracles and quantum phase
kickback for the first time in the courses under consideration and is presented in greater detail in the appendix of this paper.
\item The focus on quantum cryptography and quantum algorithms use Qiskit
\cite{Qiskit} to provide easy, hands-on access for the participants. The
code is made available to the participants, so they can also experiment
independently of the course.
\item The quantum cryptography series of courses largely dispenses with
formal mathematical formulations and uses many examples to illustrate the
concepts and mechanisms. For example, simulation examples and graphical
diagrams demonstrate entanglement swapping and purification very well.
\item In the algorithmic course series, the focus is on the practical implementation
and less on analytical investigations of the algorithms. Access to the
quantum algorithms via programming has the advantage that complex calculations
can be omitted, and a strong emphasis can be placed on demonstrations.
\end{itemize}

In addition to the MOOCs, twenty participants were selected by the HPI
to participate in a workshop associated with the courses. The workshop
was guided by members of the HPI Academy and the authors of the present
paper. It used the design thinking method (for further information see
e.g. \cite{DesignThinking}) and aimed to develop tangible applications
of quantum technologies. The workshop took the style of a business game.
A fictitious company had to be made  ``quantum-ready''
by the participants. There were no limits to the imagination. Since only
a few selected participants could attend the workshop, its evaluation is
not the topic of the present paper.

\subsection{Course Content}\label{modul_content}

The biggest challenge for the content design of the courses was the different
and heterogeneous prior knowledge of the participants. As the courses are
aimed at such an audience, only minimal mathematical and physical knowledge
could be required. On the other hand, the aim of the course series was
for participants to have a deeper understanding and overview of current
topics after successfully completing the curriculum. The bridge between
minimal prior knowledge and deeper understanding at the end of the curriculum
was built by choosing the contents of the courses as follows.%

In the introductory course series, the first course (Intro 1) introduces
logical qubits, the elementary gates X, Z, H, and CNOT, and the gate model
for the description of algorithms. Since the cube model for illustration
\cite{Just2020,Just2022} (see appendix) is used, this very quickly familiarized
the participants with quantum phenomena known from the press such as superposition
and entanglement. In the second course (Intro 2), the algorithm for teleportation
is presented. Then elementary mathematical description tools ($2^{n}
\times 2^{n}$-matrices, the operations of matrix multiplication and tensoring,
and complex numbers) are introduced. Formal definitions are deliberately
omitted. The focus is on examples and applications to give the participants
an intuitive understanding of the concepts. The third course (Intro 3)
deals with quantum oracles and the phenomenon of phase kickback, again
taking advantage of the quantum cube for visualizing abstract algorithmic
concepts. These two concepts form the basis of many classical quantum algorithms.
A brief excursion into adiabatic quantum computing concludes the introductory
course series.

In the course series on quantum cryptography, the first course (Crypto
1) explains elementary concepts from classical cryptography using examples.
The somewhat inaccessible terms computationally secure and perfectly secure
are also introduced. It is shown how quantum computers attack common asymmetric
encryption methods. The BB84 key exchange protocol \cite{Bennet} is discussed
in detail (including simple error correction and privacy amplification)
and its physical implementation and attack possibilities are explained.
In the second course (Crypto 2) the central topic is the property of entangled
quantum systems (in particular bipartite systems, Schmidt decomposition,
partial trace) and the key exchange protocol derived from this. Instead
of mathematical calculations, simulations with Qiskit \cite{Qiskit} or
quantum games (CHSH-Inequality) \cite{CHSH,CHSH_Game} are used. The ideas
on the security proofs of the protocols are also explained. The third course
(Crypto 3) discusses the concept of a quantum internet and the associated
technological challenges (repeaters and error correction).%

In the first course of the series on quantum algorithms and programming
(Algo 1), the implementation of the gate model with Qiskit
\cite{Qiskit} is presented. The classic quantum algorithms
\cite{NielsenChuang} such as teleportation, Deutsch's algorithm, Deutsch-Jozsa's
algorithm, and Grover's algorithm are presented using Qiskit. In the second
course (Algo 2), somewhat more sophisticated algorithms such as the quantum
Fourier transform, the Shor algorithm \cite{Shor} and the HHL algorithm
\cite{HHL} are explained step by step using example implementations. The
third part (Algo 3) deals with current NISQ algorithms
\cite{Preskill2018}, which can also run on existing error-prone hardware.
In addition to Monte Carlo simulations, VQE and QAOA for solving optimization
problems and algorithms are presented, see e.g. \cite{jacquier2022}.%

\subsection{Mapping to Quantum Competencies}\label{competencies}

The three Tables \ref{tabConceptFound}, \ref{tabComm} and \ref{tabAlgos} show which course teaches which competence up to which level, with the domains of possible knowledge and skills taken from the European Competence Framework for Quantum Technologies \cite{franziska_greinert_towards_2023}. The framework defines six proficiency levels: A1 Awareness, A2 Literacy, B1 Utilisation, B2 Investigation, C1 Specialisation, and C2 Innovation.

It can be seen that the curriculum covers a wide range of quantum information technology skills, mostly at level A1 of the European Competence Framework, and sometimes at level A2. This is in line with the aim of not assuming any prior knowledge, while still enabling participants to deepen their knowledge according to their needs and prior knowledge.

\begin{table}
	\centering
	\caption{Competence Domains: Concepts and Foundation}\label{tabConceptFound}
	\begin{tabular}{ |l|l|l|g|g|g|g|g|g|g|g|}
		\hline
		\rowcolor{white}\multicolumn{3}{|c|}{ } & \rot{ Intro 1 } &  \rot{ Intro 2  }  & \rot{ Intro 3 }  & \rot{ Crypto 1 }  & \rot{ Crypto 2 }  & \rot{ Crypto 3 }   & \rot{ Algo 1 }   & \rot{ Algo 2/3  }  \\
		\hline
		\rowcolor{white}  \multicolumn{3}{|l|}{Basic Quantum concepts} &  &   & &  &  &   &  &   \\
		\hline
		& \multicolumn{2}{|l|}{Superposition, Interference}& A1 &  -- &  -- & A1 & --  & --  & A1& --  \\
		\hline
		& \multicolumn{2}{|l|}{ Unitary time evolution, Schrödinger equation} & --&  A1  & A1 & --& -- & --  & A1  & --  \\
		\hline
		\rowcolor{white} & \multicolumn{2}{|l|}{Quantum measurement} & &   &  &  & & & &    \\
		\hline
		& &  Probabilistic nature of quantum physics & A1 &  -- &  -- & A1 & --  & --  & A1 & --  \\
		\hline
		& &  No cloning theorem, etc. & --&  --  & -- & A1 & -- & -- & -- & --  \\
		\hline
		\rowcolor{white} & \multicolumn{2}{|l|}{Two state systems} & &   &  &  & & & &  \\
		\hline
		& &  State representation, visualisation & A1&  A1  & A2 & A1 & A2 & --  & A1 & --  \\
		\hline
		&   \multicolumn{2}{|l|}{Pure and mixed quantum states} & -- &  --  & -- & --& A1 & --  & -- & --  \\
		\hline
		\rowcolor{white} \multicolumn{3}{|l|}{Mathematical formalism and information theory} &  &   & &  &  &   &  &   \\
		\hline
		\rowcolor{white}  & \multicolumn{2}{|l|}{Mathematical foundations} & &    &  &  & & & &  \\
		\hline
		&  & Linear algebra, functional analysis &--&  A1  & -- & A1 & -- & --  & A1  & --  \\
		\hline
		& \multicolumn{2}{|l|}{State space, Dirac notations} & A1 &  --  & -- & A1 & A1 & --  & -- & --  \\
		\hline
	\end{tabular}
 \end{table}

\begin{table}
	\centering
	\caption{Competence Domains: Quantum communication and networks}\label{tabComm}
	\begin{tabular}{ |l|p{6 cm}|g|g|g|g|g|g|g|g|}
		\hline
		\rowcolor{white}\multicolumn{2}{|c|}{ } &  \rot{ Intro 1 } &  \rot{ Intro 2  }  & \rot{ Intro 3 }  & \rot{ Crypto 1 }  & \rot{ Crypto 2 }  & \rot{ Crypto 3 }   & \rot{ Algo 1 }   & \rot{ Algo 2/3  }   \\
		\hline
		\rowcolor{white}  \multicolumn{2}{|l|}{Basics} &  &   & &  &  &   &  &   \\
		\hline
		& Conventional and PQC, combined cryptographic approaches& -- &  -- &  -- & A1 & A1 & --  &  -- & --  \\
		\hline
		& Quantum teleportation, Bell state measurement& -- &  A1 &  -- & --&  A1 & --  &  -- & --  \\
		\hline
		& Security proof, side-channel-attacks& -- &  -- &  -- & -- & A1 & --  &  -- & --  \\
		\hline
		
		\rowcolor{white}  \multicolumn{2}{|l|}{Quantum random number generators} &  &   & &  &  &   &  &   \\
		\hline
		& Secure keys, e.g. for QKD & A1&  -- &  -- & A1&  A1 & --  &  -- & --  \\
		\hline
		& Random numbers for algorithms, e.g. online gambling & A1 &  -- &  -- & -- & -- & --  &  -- & --  \\
		\hline
		
		\rowcolor{white}  \multicolumn{2}{|l|}{Quantum key distribution QKD} &  &   & &  &  &   &  &   \\
		\hline
		& QKD basic protocols, e.g. BB84, B92, E91 & -- &  -- &  -- & A1 & A2 & --  &  -- & --  \\
		\hline
		& Measurement-device independent QKD and device independent QKD & -- &  -- &  -- & --&  A1 & --  &  -- & --  \\
		\hline
		& Quantum key management systems, QKD modules (full devices) & -- &  -- &  -- & -- & -- & A1  &  -- & --  \\
		\hline
		
		\rowcolor{white}  \multicolumn{2}{|l|}{Applications of quantum cryptography} &  &   & &  &  &   &  &   \\
		\hline
		&  Secure access to cloud-based quantum computing, delegated quantum computing & -- &  -- &  -- & -- & -- & A1  &  -- & --  \\
		\hline
		
		\rowcolor{white}  \multicolumn{2}{|l|}{Quantum internet} &  &   & &  &  &   &  &   \\
		\hline
		& Quantum network nodes, memories and switches & -- &  -- &  -- & -- & -- & A1  &  -- & --  \\
		\hline
		& Quantum repeaters, entanglement swapping, entanglement purification & -- &  -- &  -- & --&  A1 & A1 &  -- & --  \\
		\hline
		&  Free-space communication& -- &  -- &  -- & A1 & -- & --  &  -- & --  \\
		\hline
		
		\rowcolor{white}  \multicolumn{2}{|l|}{Quantum internet applications} &  &   & &  &  &   &  &   \\
		\hline
		& Full quantum communication network, QKD trusted node networks (secure data transfer) & -- &  -- &  -- & -- & -- & A1 &  -- & --  \\
		\hline
		
	\end{tabular}
\end{table}

\begin{table}
	\centering
	\caption{Competence Domains: Quantum Computing and Simulation}\label{tabAlgos}
	\begin{tabular}{ |l|p{6 cm}|g|g|g|g|g|g|g|g|}
		\hline
		\rowcolor{white}\multicolumn{2}{|c|}{ } &  \rot{ Intro 1 } &  \rot{ Intro 2  }  & \rot{ Intro 3 }  & \rot{ Crypto 1 }  & \rot{ Crypto 2 }  & \rot{ Crypto 3 }   & \rot{ Algo 1 }   & \rot{ Algo 2/3  }   \\
		\hline
		\rowcolor{white}  \multicolumn{2}{|l|}{Basics} &  &   & &  &  &   &  &   \\
		\hline
		& Qubits, quantum gates, ... & A1 &  -- &  A2 &  --& -- & --  &  A1 & A2   \\
		\hline
		& Circuit design, notation, matrix representation & A1  &  A1 &  -- &  --& -- & --  &  A1 & A2 \\
		\hline
		& Basic quantum programming techniques & -- &  A1  &  -- & --& --& --  &  A1  & A2  \\
		\hline
		
		\rowcolor{white}  \multicolumn{2}{|l|}{Quantum simulators} &  &   & &  &  &   &  &   \\
		\hline
		&  Quantum annealers & -- &  -- &  A1  & --& --& --  &  -- & --  \\
		\hline
		
		\rowcolor{white}  \multicolumn{2}{|l|}{Quantum programming tools , error correction} &  &   & &  &  &   &  &   \\
		\hline
		&  Quantum assembler languages and software development kits, quantum circuit simulation	 & -- &  -- &  -- & --& --& --  &  A1 & A2 \\
		\hline
		&  Quantum compilers,high-level programming with pre-defined subroutines	 & -- &  -- &  -- & --& --& --  &  -- & A2  \\
		\hline
		&  Cloud platforms	 & -- &  -- &  -- & --& --& --  &  A1 & A2  \\
		\hline
		&  Quantum error correction	 & -- &  -- &  -- & --& -- & A1 &  -- & --  \\
		\hline
		
		\rowcolor{white}  \multicolumn{2}{|l|}{Quantum computing subroutines} &  &   & &  &  &   &  &   \\
		\hline
		&  Quantum amplitude amplification	 & -- &  -- &  -- & --& --& --  &  -- & A2  \\
		\hline
		&  QFT, hidden subgroup finding	 & -- &  -- &  -- & --& --& --  &  -- & A1  \\
		\hline
		&  Quantum phase estimation	 & -- &  -- &  A1 & --& --& --  &  -- & A2  \\
		\hline
		&  Quantum linear algebra subroutines, quantum singular value decomposition 	 & -- &  -- &  -- & --& --& --  &  -- & A1 \\
		\hline

		\rowcolor{white}  \multicolumn{2}{|l|}{Quantum algorithms} &  &   & &  &  &   &  &   \\
		\hline
		&  Numer theory and factorisation		 & -- &  -- &  -- & --& --& --  &  -- & A1  \\
		\hline
		&  Oracular algorithms and database search		 & -- &  -- &  A1& --& --& --  &  A1 & A2  \\
		\hline
		&  Linear algebra (e.g. HHL-algorithm)		 & -- &  -- &  -- & --& --& --  &  -- & A1   \\
		\hline
		&  Quantum optimisation		& -- &  -- &  -- & --& --& --  &  -- & A1  \\
		\hline
		&  Quantum machine learning, quantum neural networks  & -- &  -- &  -- & --& --& --  &  -- & A1  \\
		\hline
		&  Quantum simulation algorithms		 & -- &  -- &  -- & --& --& --  &  -- & A1  \\
		\hline
		&  NISQ-Algorithms, VQE, QAOA	 & -- &  -- &  -- & --& --& --  &  -- & A1  \\
		\hline
		
		\rowcolor{white}  \multicolumn{2}{|l|}{Application of quantum computing} &  &   & &  &  &   &  &   \\
		\hline
		&  Data security and cryptography	 & -- &  -- &  -- & --& A1& A1  &  -- & --  \\
		\hline
		
	\end{tabular}
\end{table}

\section{Evaluation of the data collected}\label{results}

To analyze the field of participants and the success of the courses, the
anonymized participant data provided by HPI is used. These data consist
of two types of data sets:
\begin{description}
\item[Dataset 1:] Once registered at OpenHPI, a pseudo ID is assigned to
each participant, which is the same for all courses attended. Now for each
course enrolment, there is an automatically generated data record with
the pseudo ID, the date of enrolment in the course, and information on
the number of course items taken (i.e. videos, self-test, and final exam),
the number of posts in the forum and the success in the final exam, if
this was taken. Information on the age of the participant is also sometimes
included if disclosed by the participant.
\item[Dataset 2:] In addition, when registering for a course, each participant
is asked to take part in a questionnaire on their personal background.
This includes questions about gender, current professional activity, and
previous education. Participation in this survey is anonymous and voluntary.
\end{description}

Across all courses in the curriculum under consideration, there are till
January 2024 a total of 17{,}157 course attendances of 7413 distinct participants.
Among those 6232 participants were already registered on OpenHPI; 1181
created a new account on the same day they enrolled in their first course.

\subsection{RQ 1: What kind of audience attended the MOOCs? What are the specific backgrounds and interests of the participants?}

In this section, the personal and professional background of the participants
is analyzed. The information provided voluntarily by the participants is
used for this purpose. Thus, the information relates to about 5700 of 17{,}147
registrations, and participants who attended several courses may have answered
more than once. Nevertheless, the analyses provide good indications of
the spectrum of participants present and of frequently occurring characteristics.

The following criteria are analyzed:
\begin{itemize}
\item[-] age (Dataset 1)
\item[-] gender (Dataset 2)
\item[-] status of employment (i.e. student, employee, pensioner) (Dataset
2)
\item[-] professional area (i.e. IT, administration, education) (Dataset
2)
\item[-] highest educational qualification (Dataset 2)
\end{itemize}

It turns out that the typical participant in a course is between 50 and
59 years old, male, employed in IT, and has a university degree. These
characteristics are mentioned in between 30\% and 80\% of registrations.
But often there is a wide variety of answers for the non-typical participants.
The courses were therefore also attended by hundreds to thousands of participants
who are female, do not work in IT, or do not have a university degree.

\begin{figure}[h]
\centering
\includegraphics[width=0.9\textwidth]{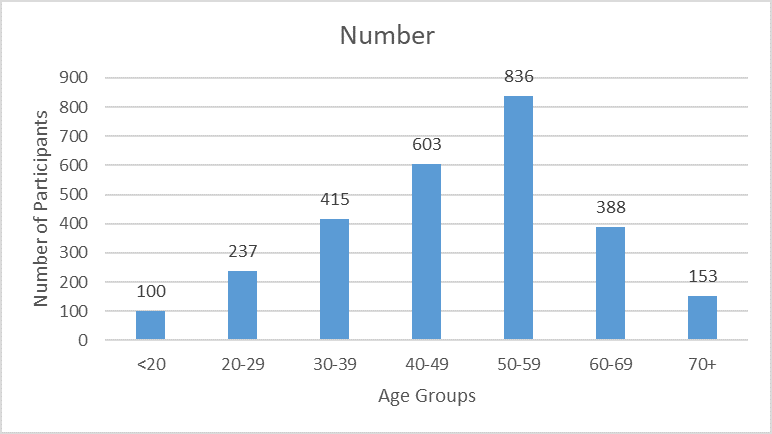}
\caption{Distribution of age groups over all participants. Participants in the survey: 2732}\label{Res_fig1}
\end{figure}

\begin{figure}
   \begin{minipage}[b]{.5\linewidth} 
      \includegraphics[width=\linewidth, height=4cm]{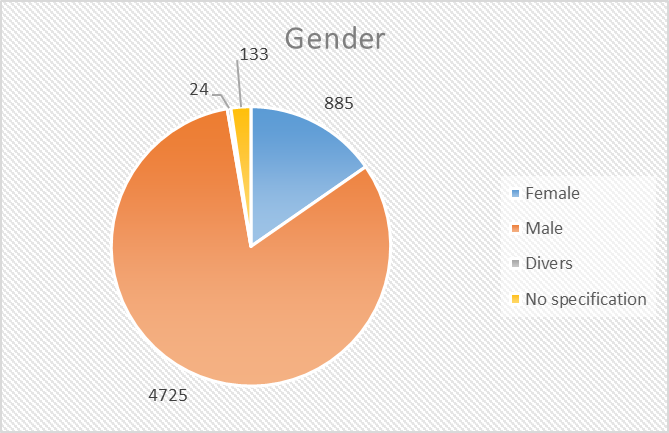}
      \caption{Distribution of different genders. Registrations in the survey: 5767}\label{Res_fig2_1_rev}
   \end{minipage}
   \hspace{.01\linewidth}
   \begin{minipage}[b]{.5\linewidth} 
      \includegraphics[width=\linewidth, height=4cm]{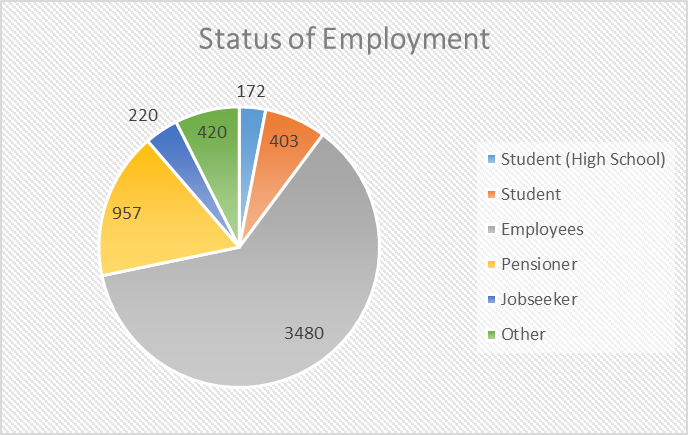}
      \caption{Staus of Employment. Registrations in the survey: 5652}\label{Res_fig2_2}
   \end{minipage}
\end{figure}

\begin{figure}
   \begin{minipage}[b]{.5\linewidth} 
      \includegraphics[width=\linewidth, height=4cm]{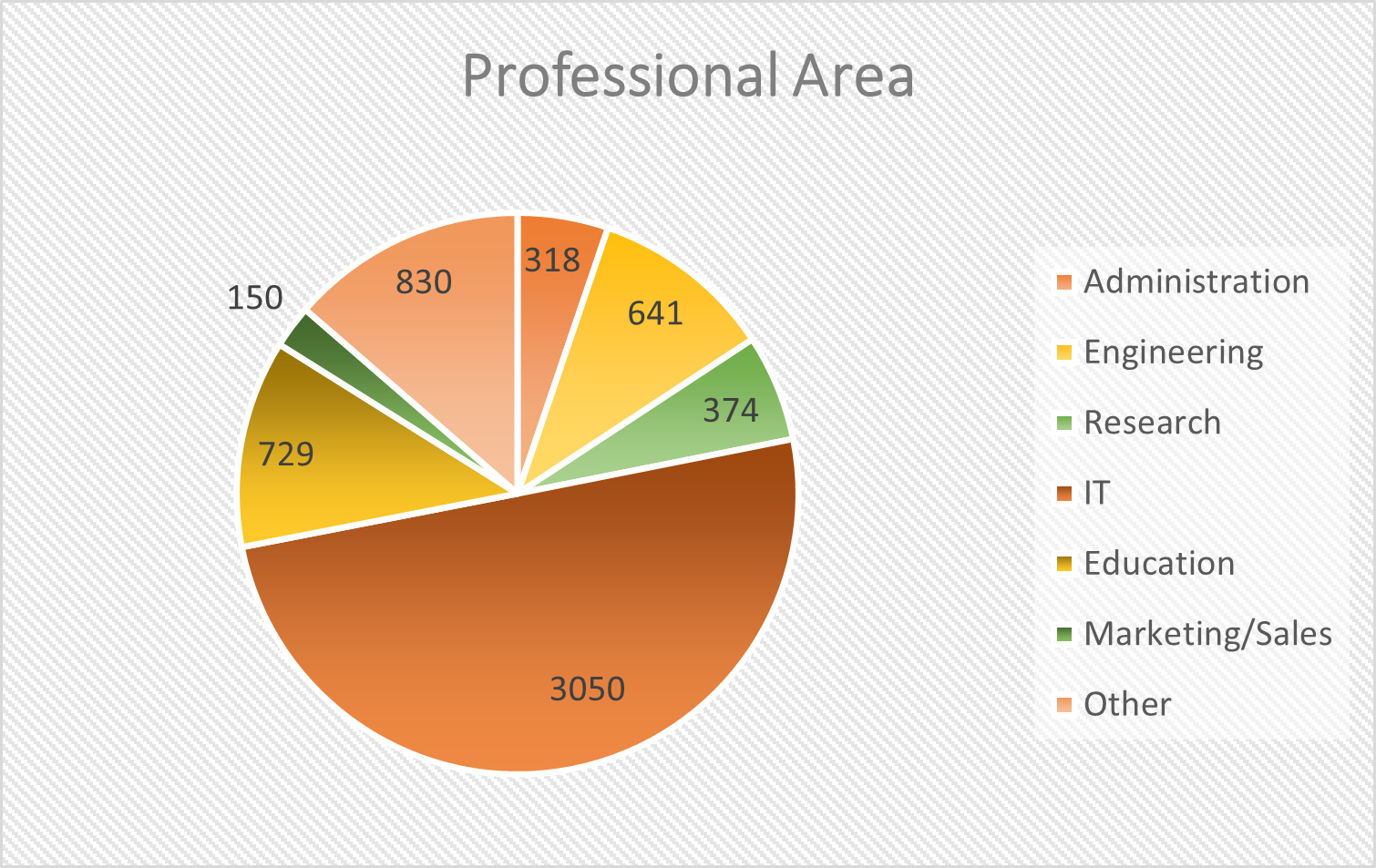}
      \caption{Professional Area of the participants. Registrations in the survey: 5303}\label{Res_fig2_3}
   \end{minipage}
   \hspace{.01\linewidth}
   \begin{minipage}[b]{.5\linewidth} 
      \includegraphics[width=\linewidth, height=4cm]{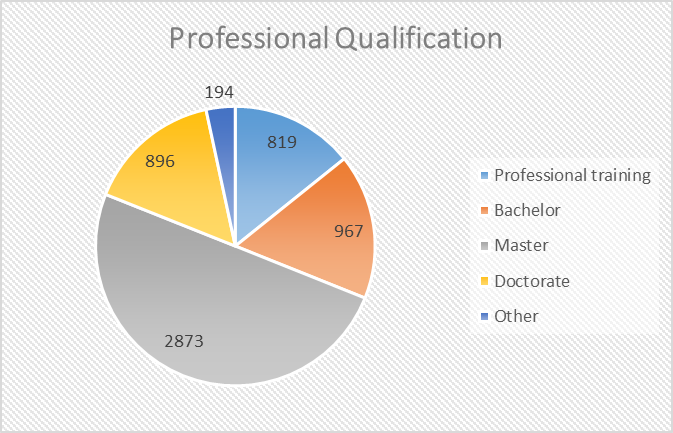}
      \caption{Professionl qualification. Registrations in the survey: 5331}\label{Res_fig2_4}
   \end{minipage}
\end{figure}

Figures~\ref{Res_fig1}-\ref{Res_fig2_4} provide detailed information:
In Fig.~\ref{Res_fig1} we show the distribution of age groups over
all participants, without double countings of participants. Unfortunately,
the majority (4747 participants) did not disclose their age. Of those who
revealed their age, a majority of participants are in the 50-59 age group.

Figure~\ref{Res_fig2_1_rev} shows the distribution of genders. At over
80\%, the proportion of men is even higher than is typically observed in
STEM subjects, where the proportion of men is regularly around 2/3.

Figures~\ref{Res_fig2_2}-\ref{Res_fig2_4} reveal information about
professional characteristics. Over 60\% of survey participants described
their employment status as employed, followed by retired people (approx.
17\%) and students at universities and others (approx. 7\% each). Unsurprisingly,
around 58\% of participants categorize themselves in the IT sector, followed
by other (around 16\%), education (around 14\%), and engineering (around
12\%). However, it was also possible to reach people from administration
and marketing/sales. In terms of professional qualifications, the MA degree
(this also includes the former German diploma degrees) dominates with approx.
54\%, followed by a Bachelor's degree (approx. 18\%) and a doctorate (approx.
17\%). Around 15\% of respondents stated that they had a non-academic professional
qualification.

\subsection{RQ 2:  What specific topics or subject areas did the participants choose when selecting the courses offered? Which topics are the participants particularly interested in?}
\label{sec4.2}

In this section, we look at the learning behavior of the participants.
The data basis is the automatically generated, anonymized data records,
i.e. there is a complete data basis without double counting (Dataset 1).
A data record refers to one participant in a course, with each participant
having an unchanging pseudo ID across all courses. The following criteria
are analyzed:
\begin{itemize}
\item[-] Number of course registrations per participant
\item[-] Number of course registrations per course
\item[-] Items visited per participant and course
\item[-] Number of registrations per learning path
\end{itemize}

It turns out that many participants specifically selected one or a few
courses and did not decide to complete the entire program with all the
courses.

Among the courses, the number of enrolments for Intro 1 is the highest,
followed by Intro 2, Crypto 1, and Algo1. The number of participants then
decreases within the three areas. This order of enrolments also holds for
the two subgroups here considered separately, namely enrolments during
the course and enrolments for self-study after the course was finished.

When analyzing the items attended per course, after deducting the no-shows
who were only enrolled and did not attend any items at all, the largest
group of participants attended 75\% to 100\% of the items. The second-largest group only attended less than 25\% of the items. It turns
out that only a small percentage of participants visited an average number
of items between 26\% and 75\%.

When analyzing the learning paths, only participants who attended at least
25\% of the items in the courses are considered, as only here can the effect
of the didactic concept unfold. It turns out that the majority of these
participants followed the recommended learning paths.

\begin{figure}[h]
\centering
\includegraphics[width=0.9\textwidth]{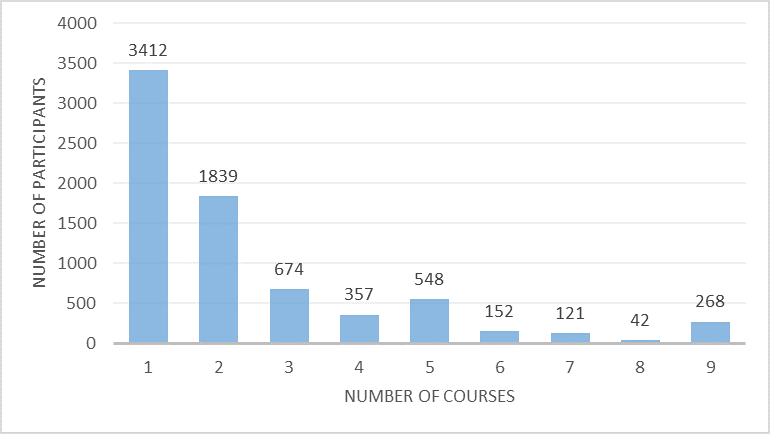}
\caption{Number of participants taken one, two, ..., nine courses.}\label{Res_fig3}
\end{figure}

\begin{figure}[h]
\centering
\includegraphics[width=0.9\textwidth]{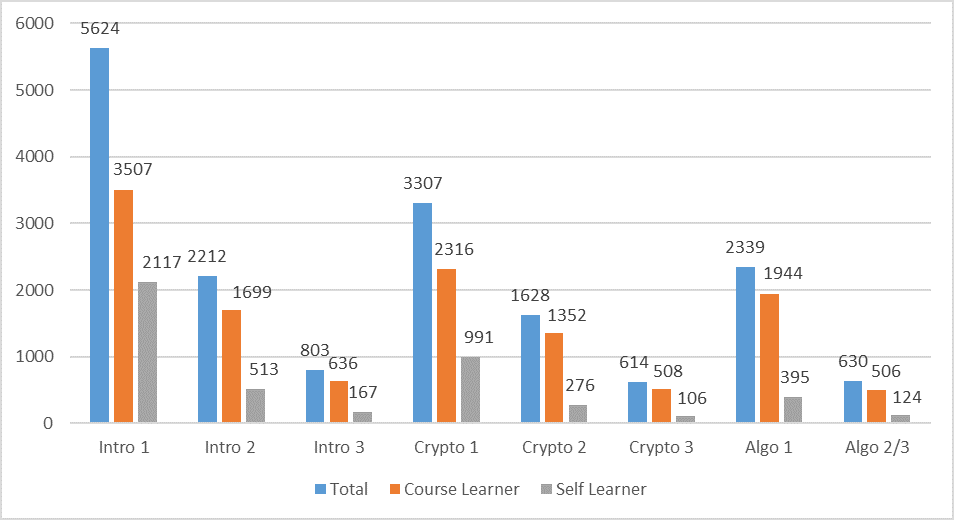}
\caption{Registrations per course, subdivided to course learners and self-learners.}\label{Registrierung_CL_SL}
\end{figure}

\begin{figure}[h]
\centering
\includegraphics[width=0.9\textwidth]{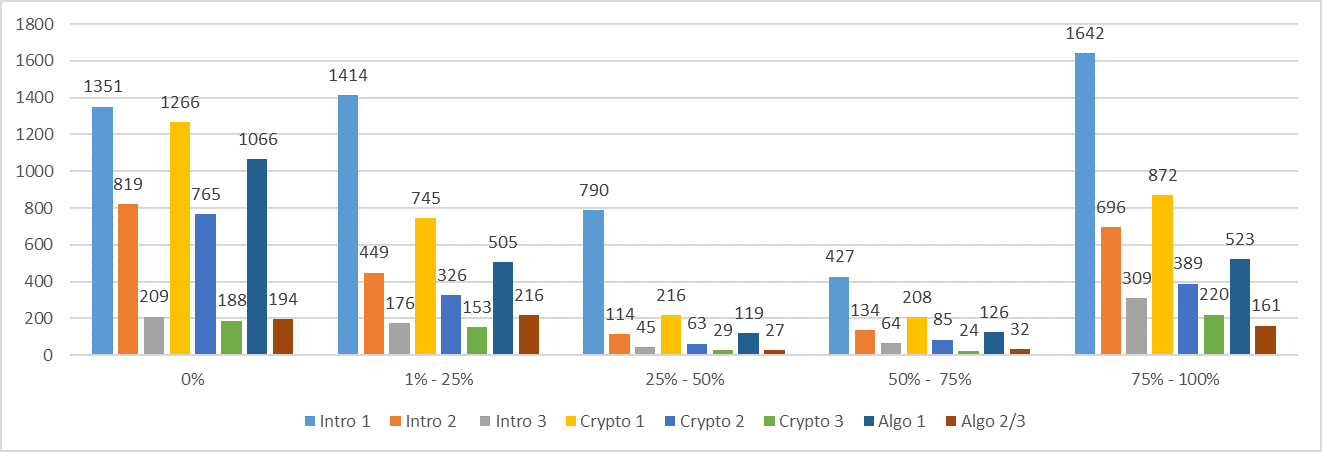}
\caption{Number of participants using percentage of the items.}\label{ItemsVisited}
\end{figure}

\begin{figure}[h]
\centering
\includegraphics[width=10cm]{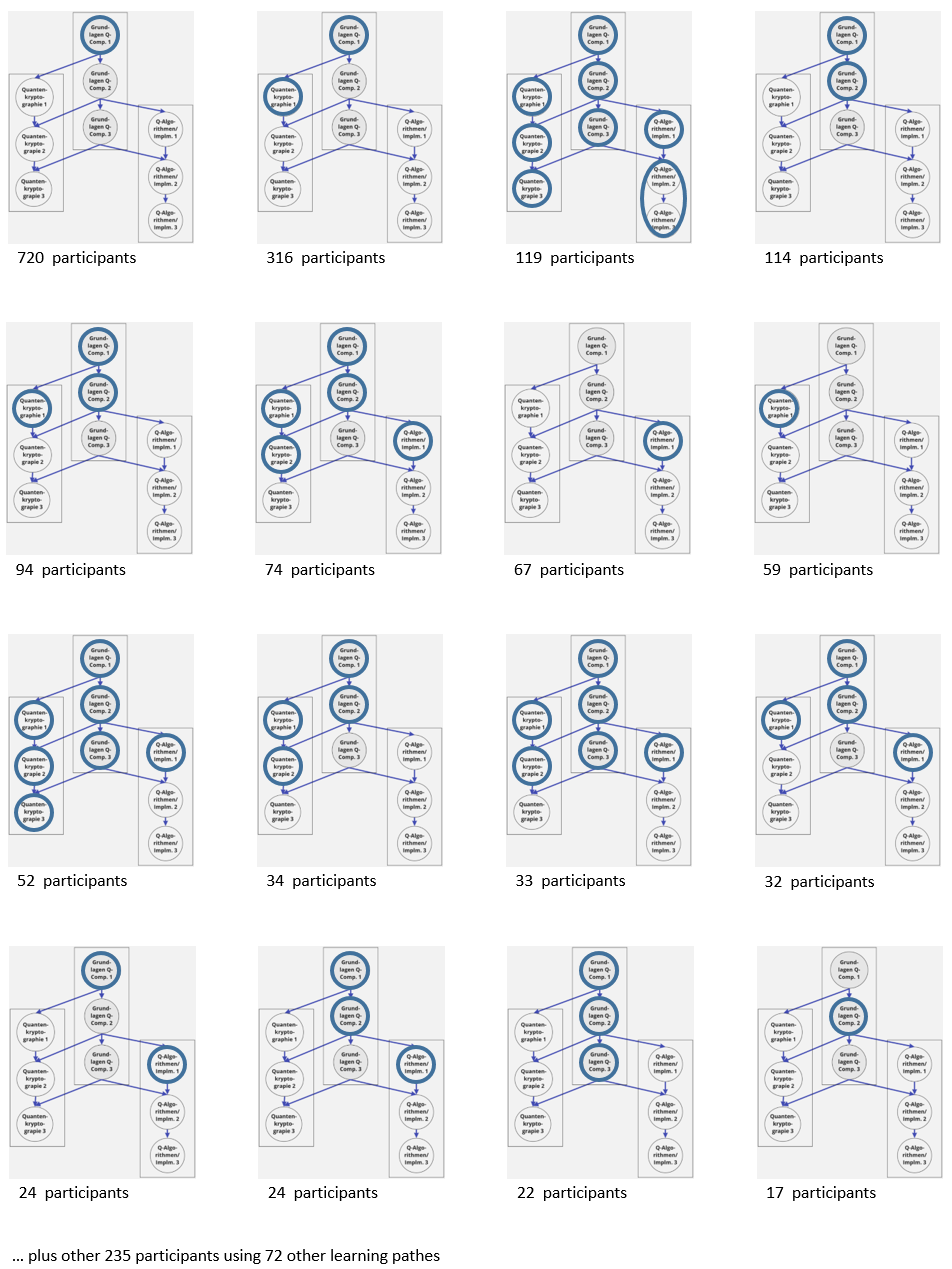}
\caption{The 16 most frequently used learning paths embedded to the untranslated German curriculum.}\label{Lernpfade}
\end{figure}

Figures~\ref{Res_fig3}-\ref{Lernpfade} show the details. Figure~\ref{Res_fig3}
shows that many participants specifically selected one or a few courses
and did not decide to complete the entire program with all courses. The
number of participants enrolling in one course, two courses, ..., and all
nine courses can be seen.

Figure~\ref{Registrierung_CL_SL} shows the registrations per course.
Summed up over all nine courses, there are 17{,}157 registrations until January
2024. 12{,}468 are course learners and 4689 are self-learners. The courses
continue to be open for self-learners, so the number of self-learners is
still increasing.

Figure~\ref{ItemsVisited} displays for each course initially how many
participants visited how many items. Participants are divided into 5 groups:
the no-shows (participants who registered but did not visit any course
item), and then the 4 groups of participants who visited between 1\% and
25\%, between 26\% and 50\%, between 51\% and 75\%, and more than 75\%
of the items. In all courses, the middle participant groups, those with
shares between 26\% and 75\%, were by far the smallest groups. The other
groups are roughly equal in size, with variations between individual courses.

Figure~\ref{Lernpfade}  shows which participants took which learning
paths. Here, only course participations with at least 25\% of visited items
are considered, these are 2036 participants. The most visited learning
path consists of the course Intro 1 only, with 720 participants. This is
followed by the learning path Intro 1 and Crypto 1 with 316 participants,
and then the learning path with all courses with 119 participants. Only
in exceptional cases course selections outside   the recommended
learning paths were made.


\subsection{RQ 3: Which competencies could be acquired by the participants through the course concept? To what extent did the course concept support the participants?}
\label{sec4.3}

To check the success of course participation, we split the 17{,}157 data records
into 12{,}468 data records from course learners (CLs) and 4689 data records
from self-learners (SLs). The distinction between course learners and self-learners
was made in the data records based on the enrolment date for the course.%

Course learners used the same learning items as self-learners, except that
they also had the opportunity to exchange ideas and ask questions in the
forum. They also took a final exam.

The learning success of course learners is derived from the result of the
final examination. This is passed if 50\% of the questions are answered
correctly. These participants get a  ``Record of Achievement''.
All participants who visited at least 50\% of the items in a course get
a  ``Confirmation of Participation''.%

It can be seen that of the 12{,}468 course learners, 4550 are no-shows, 3145
did not reach the 50\% items threshold and 4773 successfully completed
the course. This corresponds to 38\% of all course learners.

\begin{figure}[h]
\centering
\includegraphics[width=0.9\textwidth]{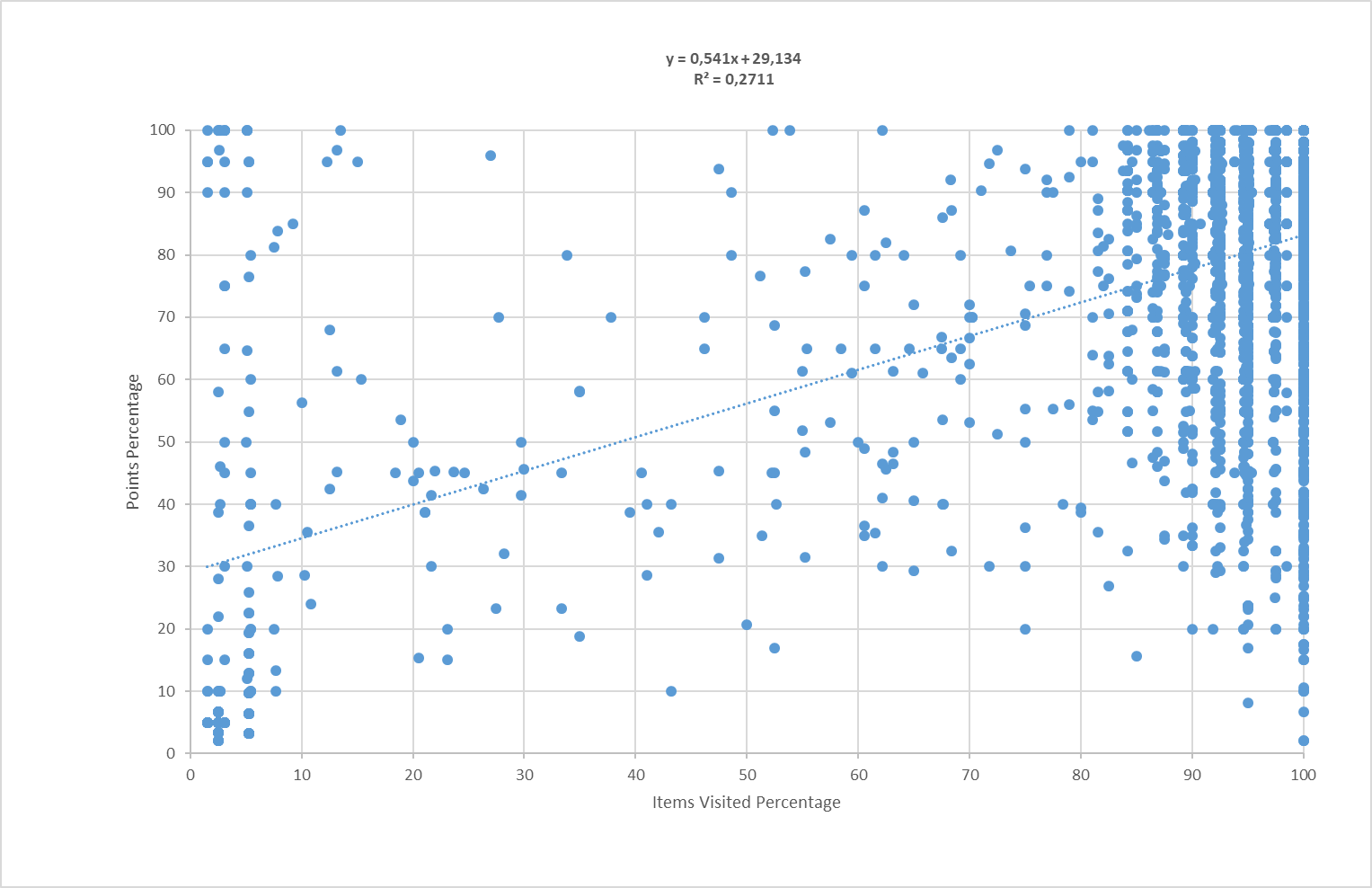}
\caption{Points percentage (=total score) of the course learners vs. percentage of items visited. 3299 data points are included.}\label{Res_fig4}
\end{figure}

To gain more insights on which the points percentage (total score) of the course learners may depend, we model this with a simple regression model where the points percentage depends linearly on the percentage of items visited, see
Fig.~\ref{Res_fig4}. Here, besides the data points, the linear model with slope and intercept as well as the $R^{2}$ can be seen.
 While only a few participants achieved a high overall score despite having
attended fewer learning items, the majority showed a clear positive correlation
between the two variables
as seen by the value of $R^{2}=0.27$. Note, that as a global model, the linear regression interpolates smoothly between different regimes of the data.

Aggregated over all courses, 3299 participants tried the final exams and
are included in this analysis. This underlines the significance of the
learning materials visited for course success.

\begin{figure}[h]
\centering
\includegraphics[width=0.9\textwidth]{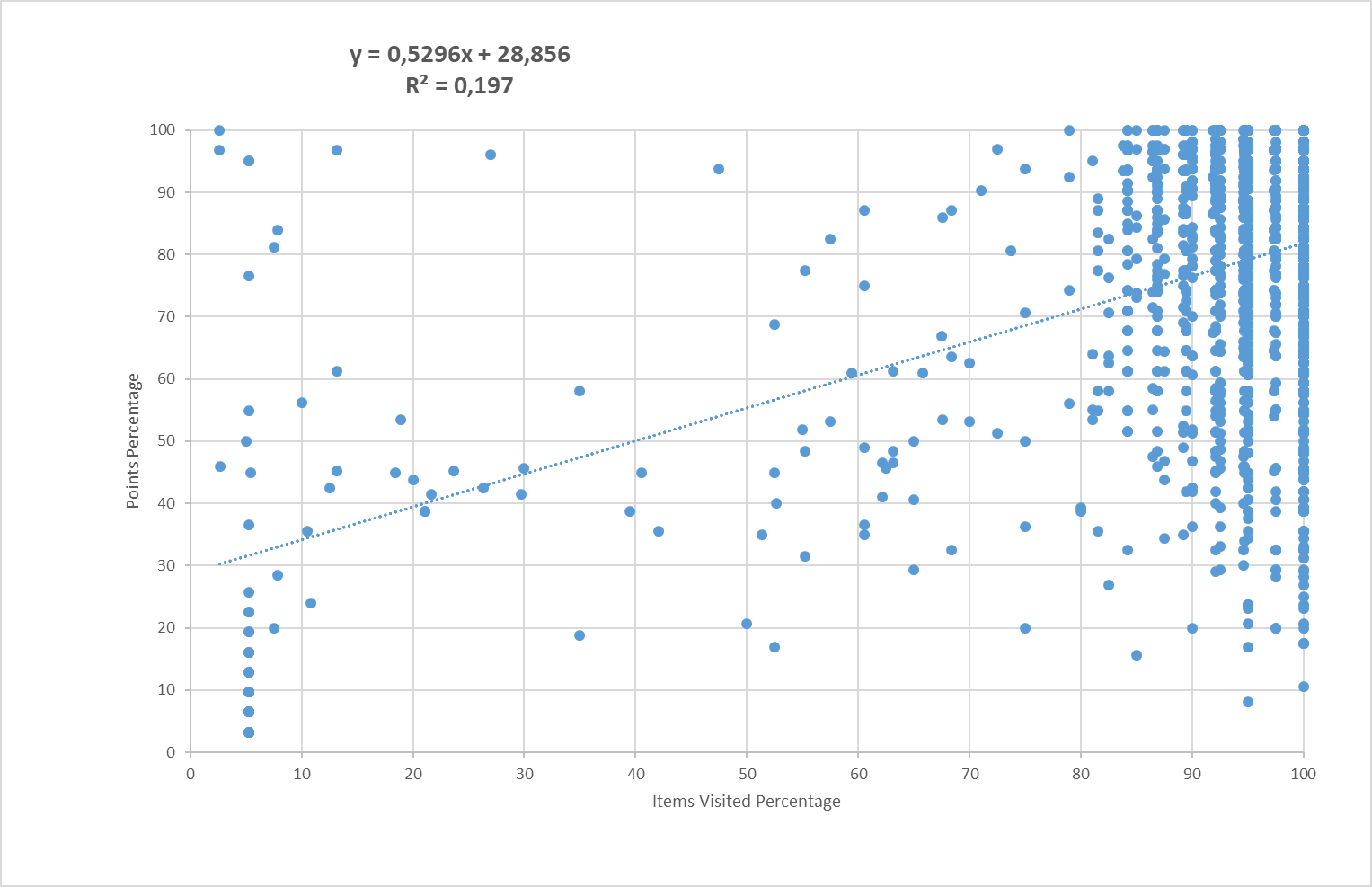}
\caption{Courses Intro 1, Intro 2 and Intro 3: Points percentage (=total score) of the course learners vs. percentage of items visited. 1667 data points are included.}\label{Res_fig4_Intro}
\end{figure}

\begin{figure}[h]
\centering
\includegraphics[width=0.9\textwidth]{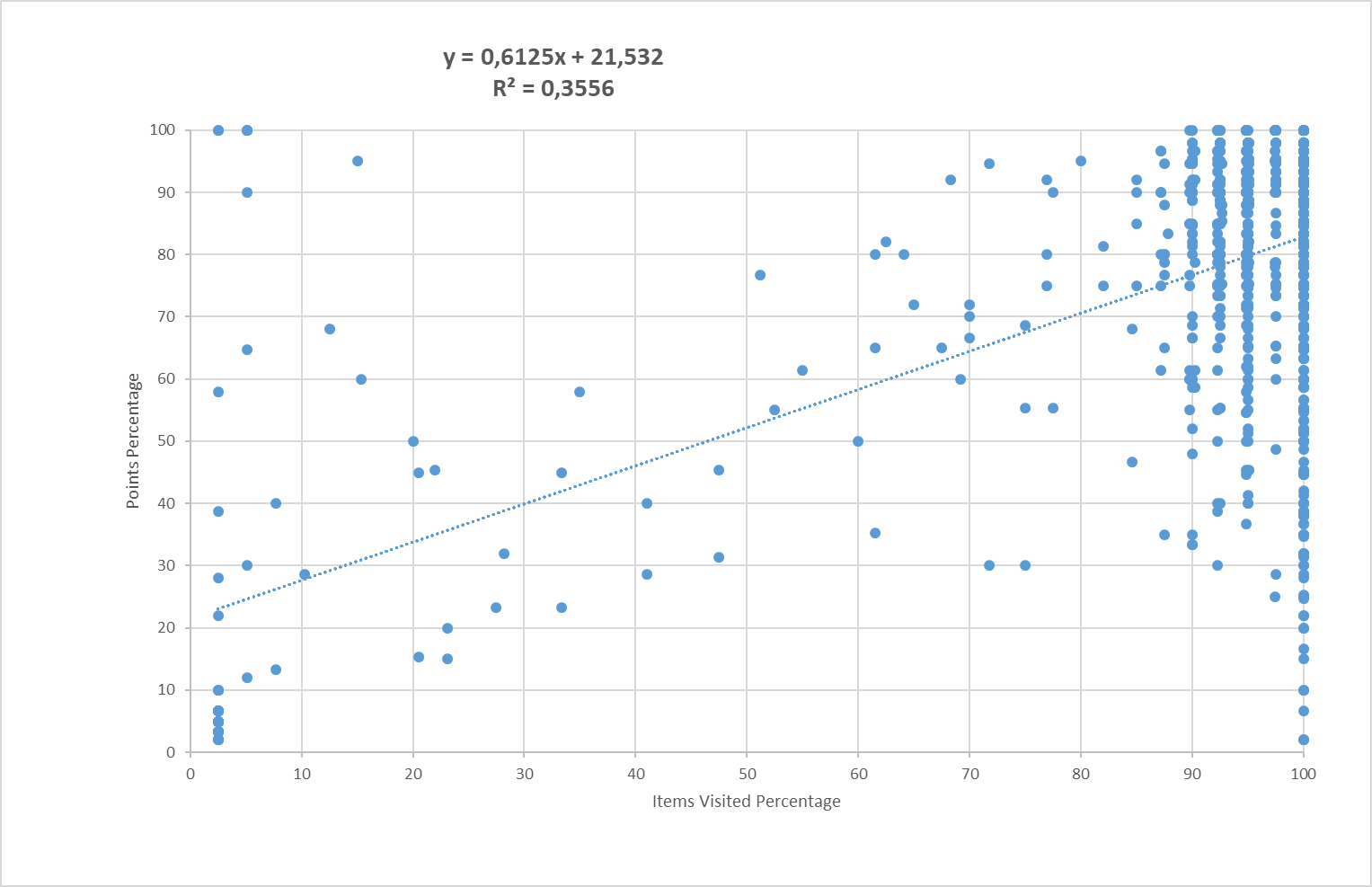}
\caption{Courses Crypto 1, Crypto 2 and Crypto 3: Points percentage (=total score) of the course learners vs. percentage of items visited. 1051 data points are included.}\label{Res_fig4_Crypto}
\end{figure}

\begin{figure}[h]
\centering
\includegraphics[width=0.9\textwidth]{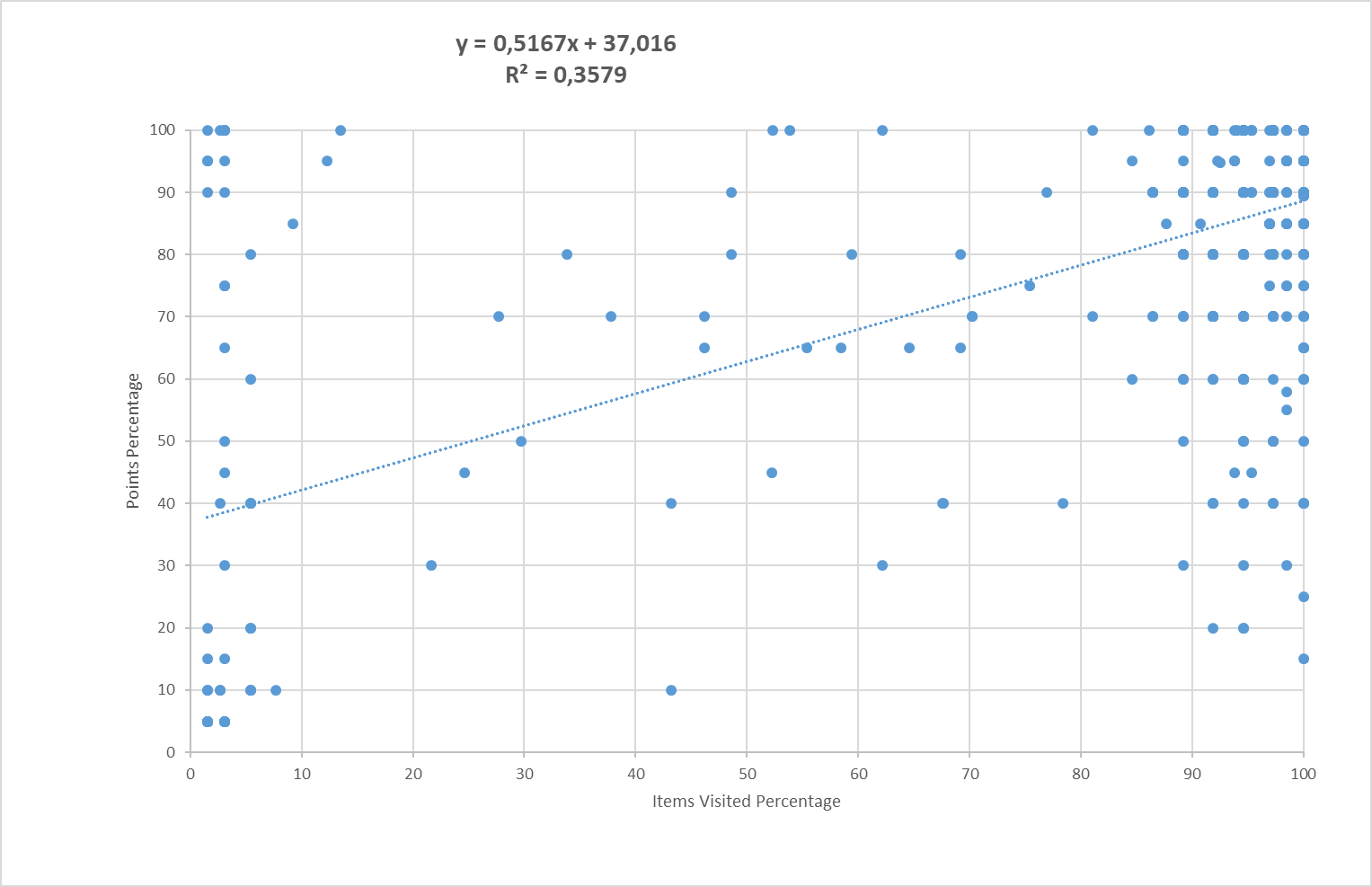}
\caption{Courses Algo 1 and Algo 2/3: Points percentage (=total score) of the course learners vs. percentage of items visited. 581 data points are included.}\label{Res_fig4_Algo}
\end{figure}

When dividing the dataset into the three course blocks, Intro 1 to 3, Crypto
1 to 3, and Algorithms 1 to 2/3, we observe a qualitatively similar
dependence on points percentage and items visited percentage: Participants
using the content of the courses more actively performed better in the
final exam. This is shown in Figs.~\ref{Res_fig4_Intro}  to  \ref{Res_fig4_Algo}.

\begin{figure}[h]
\centering
\includegraphics[width=0.9\textwidth]{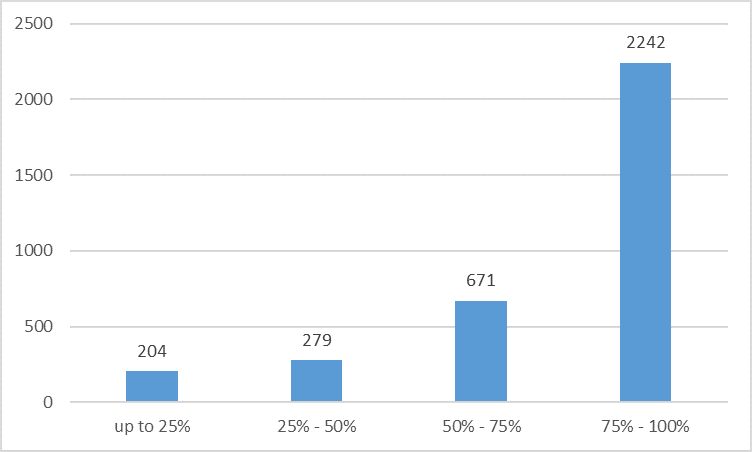}
\caption{Number of participants achieving points percentage of the final examination.}\label{DistributionResults}
\end{figure}

\begin{figure}[h]
\centering
\includegraphics[width=0.9\textwidth]{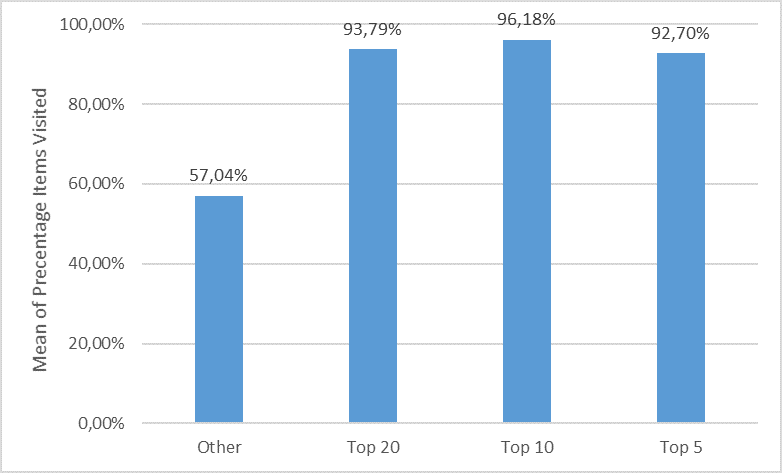}
\caption{Percentage of the items visited by Top Performers (Top 20, Top 10, Top 5) and the other course-learners.}\label{Res_fig5}
\end{figure}

For the course learners, we can furthermore relate the success
of the final exam to the items visited. Figure~\ref{DistributionResults}
gives an overview on the results achieved in the final exams. In Fig.~\ref{Res_fig5} we show that the mean of the number of items visited is
significantly larger for the top performers of the courses (top 5\%, top
10\% and top 20\% of the participants) compared with the remaining participants.
Here  ``top\textit{x}'' means that a participant has achieved
one of the \textit{x} highest percentage results.
Surprisingly, the Top 5 performers attended slightly fewer items than the Top 10 and Top 20 performers. This may be due to the fact that some quantum professionals also attended the courses to relate them to their professional knowledge or to use the structure of the courses. Overall, the
 top performers visited more than 92\% of the items on average.

It can be seen that of the 4689 self-learners, 1369 were no-shows, 2181
did not visit 50\% of the items and 1139 finished the course with success,
so roughly a portion of 24\% of the participants starting the course completed
it with a certificate of attendance.

\begin{figure}[h]
\centering
\includegraphics[width=0.7\textwidth]{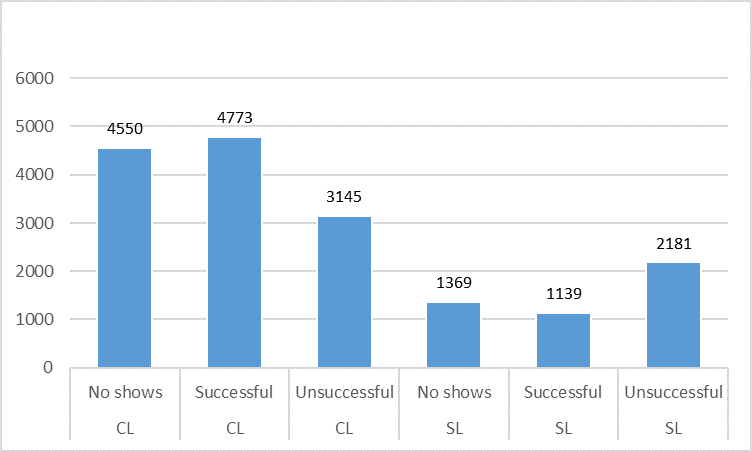}
\caption{Divided between course and self-learners the number of no-shows, successful and unsuccessful participants are displayed.}\label{SuccessfulParticipants}
\end{figure}

Figure~\ref{SuccessfulParticipants} gives an overview and the details
on course registrations for course learners and self-learners, the no-shows,
the unsuccessful, and the successful participants. Across all courses and
participants, 5912 completed the course; in 11{,}245 cases the courses were
not completed. It should be noted here that individual participants are
counted separately in each course they attended.

\section{Conclusions and topics for further research}\label{conclusion}

In the previous section, we analyzed the participant's data and the data
of their learning behavior in the core courses on quantum computing on
the OpenHPI platform that took place in the period from June 2022 to July
2023. This provides answers to the following research questions
\begin{enumerate}[RQ 3:]
\item[RQ 1:] What kind of audience attended the MOOCs? What are
the specific backgrounds and interests of the participants?
\item[RQ 2:]
 What specific topics or subject areas did the participants choose when selecting the courses offered?
Which topics are the participants particularly interested in?
\item[RQ 3:] Which competencies could be acquired by the participants
through the course concept? To what extent did the course concept support
the participants?
\end{enumerate}
It should be emphasized once again that all data included in the analyses
is anonymized so that no conclusions can be drawn about individual participants.

RQ1: Our analyses show that the quantum MOOCs were attended by
an audience with different backgrounds and interests. All age groups were
represented, with the majority being over forty. Most participants had
an IT background, but participants from non-technical professions are also
a considerable part of the audience. The majority of participants
  have a university degree. As is often the case in
technical and scientific fields, male participants also predominated in
our courses. The analyses show that people not directly associated with
a university are also interested in quantum computing. An enlarged
audience with different professional and personal characteristics could
therefore be addressed.

RQ2:  As the field of quantum computing is very broad, various
learning paths were suggested to the participants. With the MOOC platform's
help, the participants' learning behavior could be tracked in detail. The
analyses show that the recommended learning paths were often followed.
Most participants attended an introductory course in each subject area
(Intro 1, Crypto 1, Algo 1). Depending on their interests, they remained
loyal to the topic and completed more advanced courses. A not inconsiderable
number of participants also attended courses on several topics. More than
140 people completed all nine core MOOCs. The data shows that the recommendation
of learning paths was highly accepted. Thus, there seems to be evidence
that it is important to suggest learning paths to participants for better
orientation when dealing with complex topics.

RQ3:  The generally recognized European Competence Framework for
Quantum Technologies \cite{franziska_greinert_towards_2023} was used to
select material. All content taught and tested in the examinations could
be assigned to categories in the reference framework. Over 80\% of exam
participants passed the final exam, 50\% even with a good to excellent
result. It also shows that the success rate correlates with the number
of learning materials consumed (videos, self-study tests, forum discussions).
It can be stated that the learning elements provided by the MOOC platform
are sufficient to teach participants the necessary skills.

In conclusion, one can say that the quantum MOOCs at OpenHPI were a big
success and up to January 2024 reached more than 7400 participants from
different age and professional groups. The MOOCs help to bring the topic
of quantum computing closer to an audience outside of universities
and research departments. The MOOCs are a good opportunity for anyone interested
in quantum computing to delve deeper into the topics. The structure and
content of the courses might also be a good point of reference and starting
point for persons who would like to hold quantum computing courses. Depending
on the desired learning objectives, teachers can build their own courses
along the proposed learning paths, since the learning paths provide a self-contained
arrangement of quantum computing topics. The teachers then can
deepen quantum topics according to their specialization, and insert additional
topics according to their preferences.

As the courses are in high demand, the openHPI has now started to provide the videos with English subtitles to make them accessible to a non-German-speaking audience. However, as the slides shown are still in German, the reach of the courses is likely to remain limited. In hindsight, it would have been better to formulate the slide content directly in English.
The project schedule (start, milestones, end) required by the project organizer meant that the timing of the courses was not always optimal. Some courses were held very close together, while other courses were sometimes several months apart. When organizing an extensive series of courses, care should be taken to ensure that the courses are evenly distributed, giving the participants enough time to catch up on their work, but that the breaks between the courses are not too long.

The experience gained from running the MOOCs shows that many concepts of
quantum computing can be explained to an interested audience without complex
mathematical and physical descriptions. Presenting the various topics using
examples and applications that are closer to everyday life has proved useful.
Mathematical descriptions were only used when there was an intuitive understanding
based on the application. Such an approach lowers the entry barrier to
the topic and leads to less frustration among learners. Mathematical and
physical descriptions and concepts should only be introduced and used once
a basic understanding and curiosity have been aroused.

In addition to the results discussed in the present paper, further research
questions can be investigated based on the available data. It would be
interesting to see, how exactly the participants worked with the video
tutorials. For example, were the videos paused in between to better understand
complicated details? Did the sometimes detailed discussions in the forum
motivate the participants to engage with the topic? Another interesting
question could deal with possible specific learning differences between
male and female participants. More generally, the learning behavior of
different subgroups could be analyzed in relation to age, professional
situation, or educational attainment.
The professional and personal backgrounds of the Top Performers could also be analyzed in more detail.
Since the amount of data is large, the subgroups usually consist of hundreds
of participants with respect to various characteristics, so that statistical
analyses are meaningful.

\backmatter


\bmhead{Acknowledgements}
The authors would like to thank the Hasso-Plattner-Institut f\"{u}r Digital Engineering gGmbH and in particular Christoph Meinel and Martin van Elten for providing the data on which this study is based, and the unknown reviewers for their valuable hints.

\section*{Declarations}

\begin{itemize}
\item Funding:
The work of Gerhard Hellstern is partly funded by the Ministry of Economic Affairs, Labour and Tourism Baden-Württemberg in the frame of the Competence Center Quantum Computing Baden-Württemberg (project \lq QORA II\rq).

\item Conflict of interest/Competing interests: 
The authors have no relevant financial or non-financial interests to disclose.

\item Ethics approval and consent to participate:
Not applicable. 

\item Consent for publication:
Not applicable.

\item Data availability:
The data that support the findings of this study are available upon reasonable request.

\item Materials availability:
Not applicable.

\item Code availability:
Not applicable.

\item Author contribution:
The authors contributed equally to the paper. All authors read and approved the final manuscript.
\end{itemize}

\begin{appendices}
\section{The ``Quantum Cube" for visualizing entanglement in multi-qubit-registers and the effect of quantum gates}

In the introductory MOOCs of the curriculum, the  ``quantum cube''
is used throughout to illustrate quantum register states and the effect
of quantum gates.

\begin{figure}[h]
\centering
\includegraphics[width=0.5\textwidth]{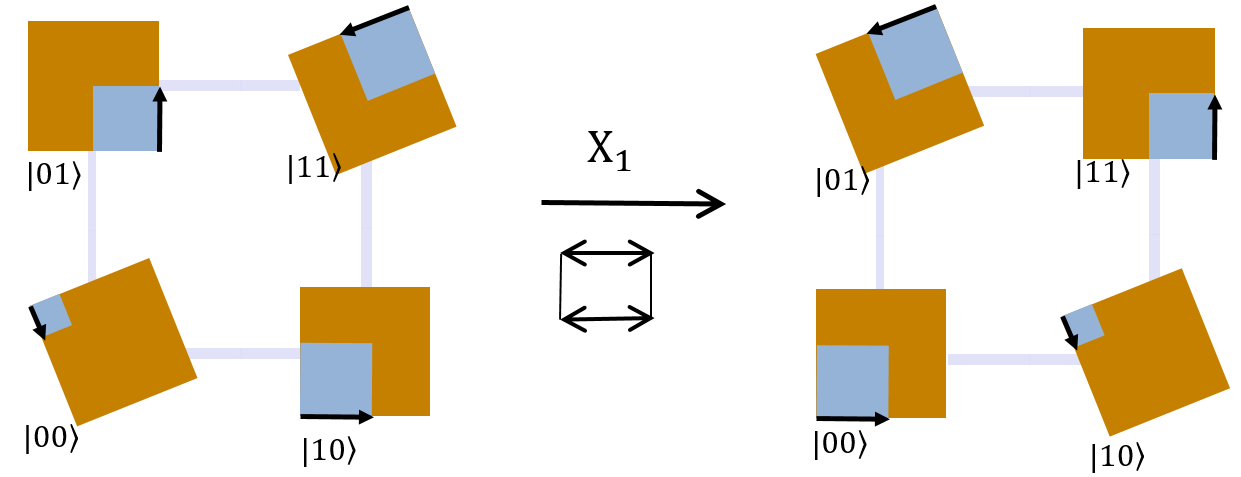}
\caption{Visualizing the state $\sqrt{\frac1{16}}e^{i290^o}|00\rangle + \sqrt{\frac4{16}}i|01\rangle + \sqrt{\frac5{16}}|10\rangle + \sqrt{\frac6{16}}e^{i200^o}|11\rangle$ of a 2-qubit-register and the effect of Pauli-X operating on the first qubit with the quantum cube model. The first qubit is provided with the left-right direction in 2-dimensional space, the second qubit with the down-up direction. The amplitudes are placed at the corresponding corners of the n-dimensional unit cube, which in 2 dimensions is the unit square. The representation of the complex amplitudes here follows Feynman's QED \cite{Feynman}. Pauli-X on the first qubit interchanges the amplitudes along the left-right lines of the ``cube". The resulting state is 
$\sqrt{\frac5{16}}|00\rangle + \sqrt{\frac6{16}}e^{i200^o}|01\rangle + \sqrt{\frac1{16}}e^{i290^o}|10\rangle + \sqrt{\frac4{16}}i|11\rangle$.} \label{CubeModel1}
\end{figure}

The model was developed by Just \cite{Just2020,Just2022} and since then
has been studied and used as an intuitive tool for visualizing quantum
entanglement and quantum computation
\cite{seegerer2021quantum,Fraunhofer2022,bley2024visualizing}. The quantum
cube represents the state of an n-qubit register by providing one dimension
in n-dimensional Euclidean space to each qubit. It then utilizes the standard
n-dimensional cube with corners $(0,\ldots,0)$ up to $(1,\ldots,1)$. For a given
register state, it places the amplitudes of the basis states (in any suitable
representation of numbers, i.e. digits, arrows, squares, cubes, or circles)
at the corresponding corners of the n-dimensional cube. See Fig.~\ref{CubeModel1}
for an example with two qubits, and the effect of Pauli-X-gate operating
on the first of them.

\begin{figure}[h]
\centering
\includegraphics[width=1.0\textwidth]{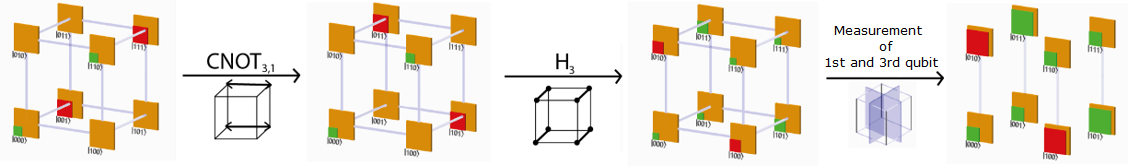}
\caption{The CNOT-Hadamard-Measurement-steps of the teleportation algorithm as in  \cite{Just2020}. For didactical reasons, only real amplitudes are considered. Green squares indicate positive amplitudes, red squares indicate negative amplitudes. The state of a register with three qubits is visualized in a 3-dimensional cube. Each qubit is provided with one dimension in space: The first qubit with the left-right dimension, the second qubit with the down-up direction, the third qubit with the front-back direction. }\label{CubeModel2}
\end{figure}

The application of quantum gates to a quantum register can
generally   be visualized as a swapping, rotation,
or mixing of the numbers at the corners of the cube. This has been done
by Just \cite{Just2020,Just2022} for the teleportation algorithm, see
Fig.~\ref{CubeModel2}.

\begin{figure}[h]
\centering
\includegraphics[width=0.7\textwidth]{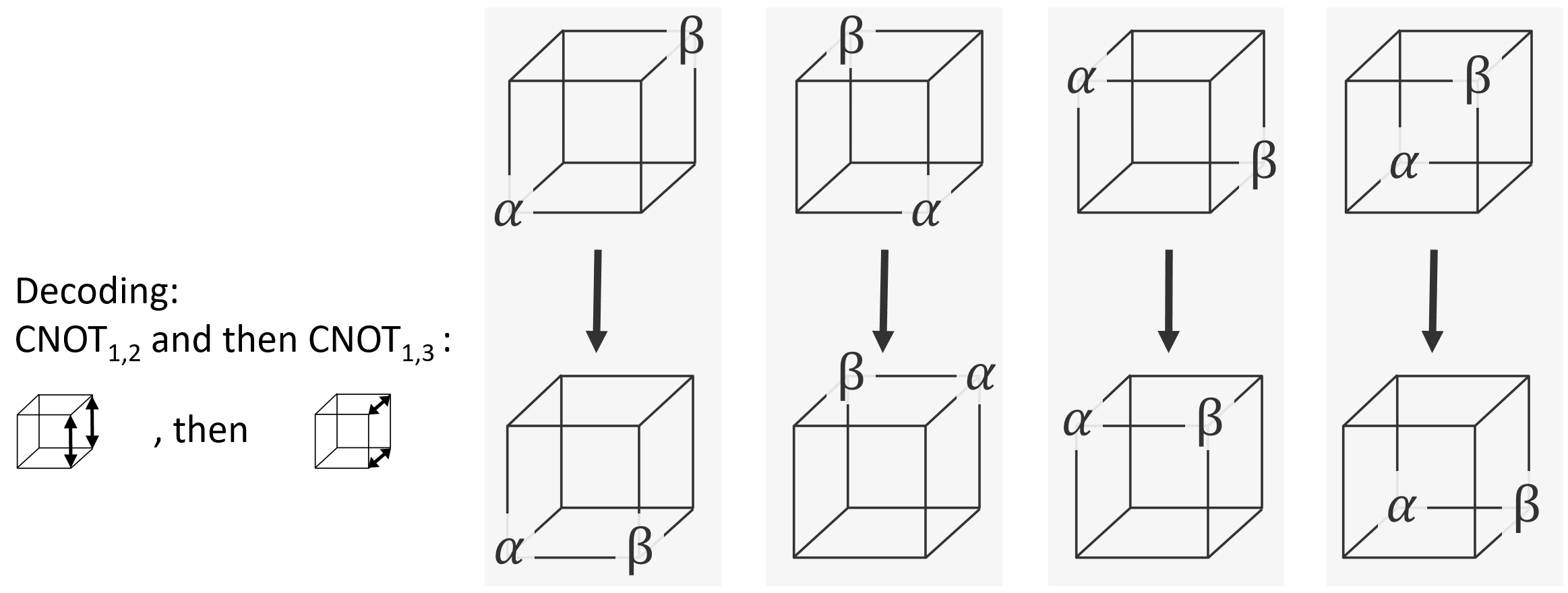}
\caption{The decoding in bit flip correction code, visualised with the quantum cube model. The CNOT gate with the first qubit as control-bit is applied two times, first with the second qubit as target-bit (down-up interchangement of amplitudes when first qubit equals $|1\rangle$), then with the third qubit as target-bit (front-back interchangement of amplitudes when first qubit equals $|1\rangle$). It can be seen how the amplitudes $\alpha$ and $\beta$ of the first qubit (left-right direction) are rearranged to one specific state of the syndrome qubits (down-up and front-back direction).}\label{CubeModel3}
\end{figure}

\begin{figure}[h]
\centering
\includegraphics[width=0.7\textwidth]{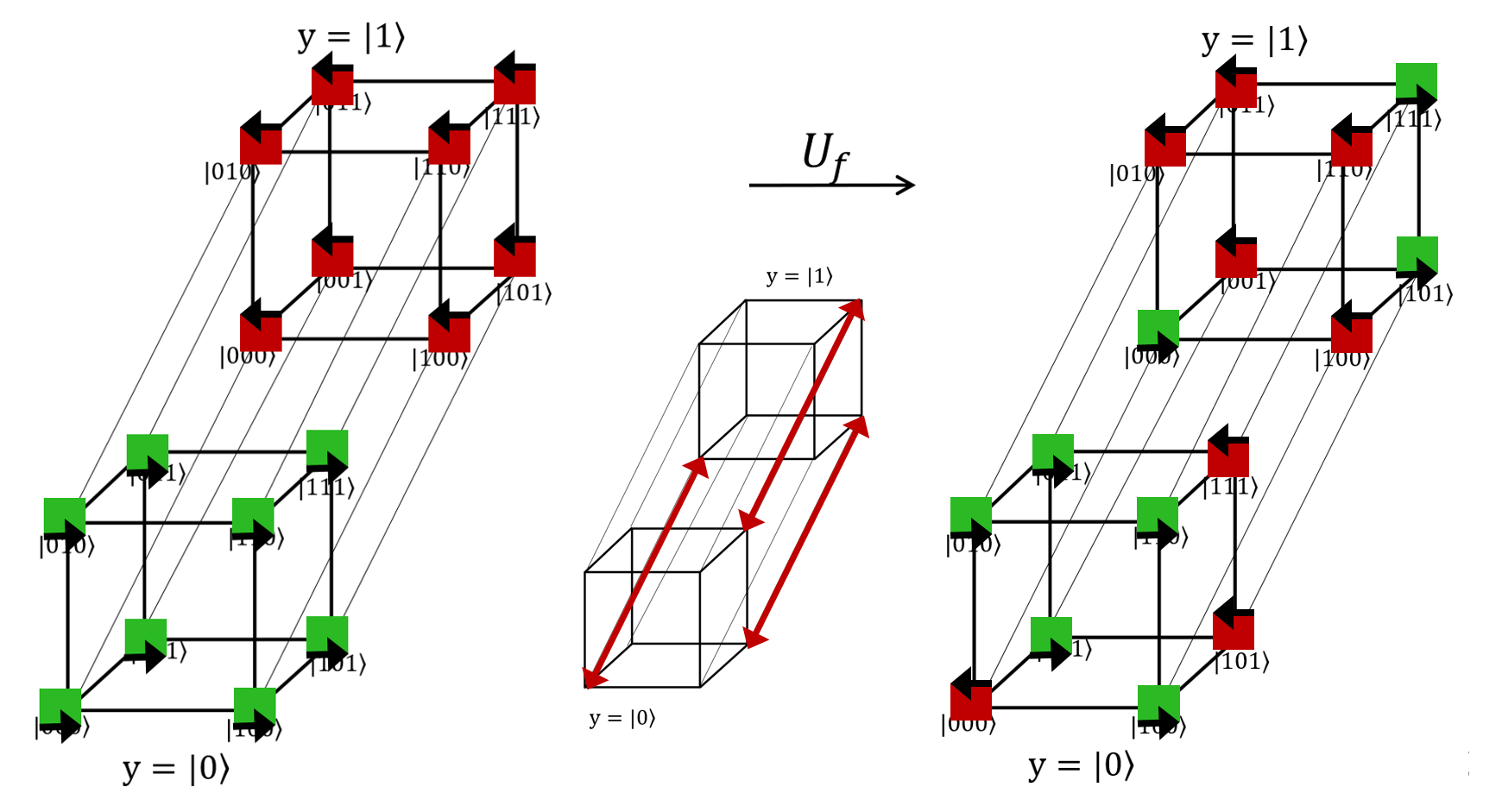}
\caption{A quantum oracle for a boolean function of three bits, in this example $f(x_1, x_2, x_3)=1$ if and only if $(x_1, x_2, x_3)$ is $(0,0,0)$ or $(1,0,1)$ or $(1,1,1)$. The oracle $U_f$ in general interchanges the amplitudes of the $|x_0\rangle$ and $|x_1\rangle$ states, if and only if $f(x)=1$. The initial state in this figure is prepared as it is done in the algorithms of Deutsch-Jozsa and Bernstein-Vazirani}\label{CubeModel4}
\end{figure}

\begin{figure}[h]
\centering
\includegraphics[width=0.9\textwidth]{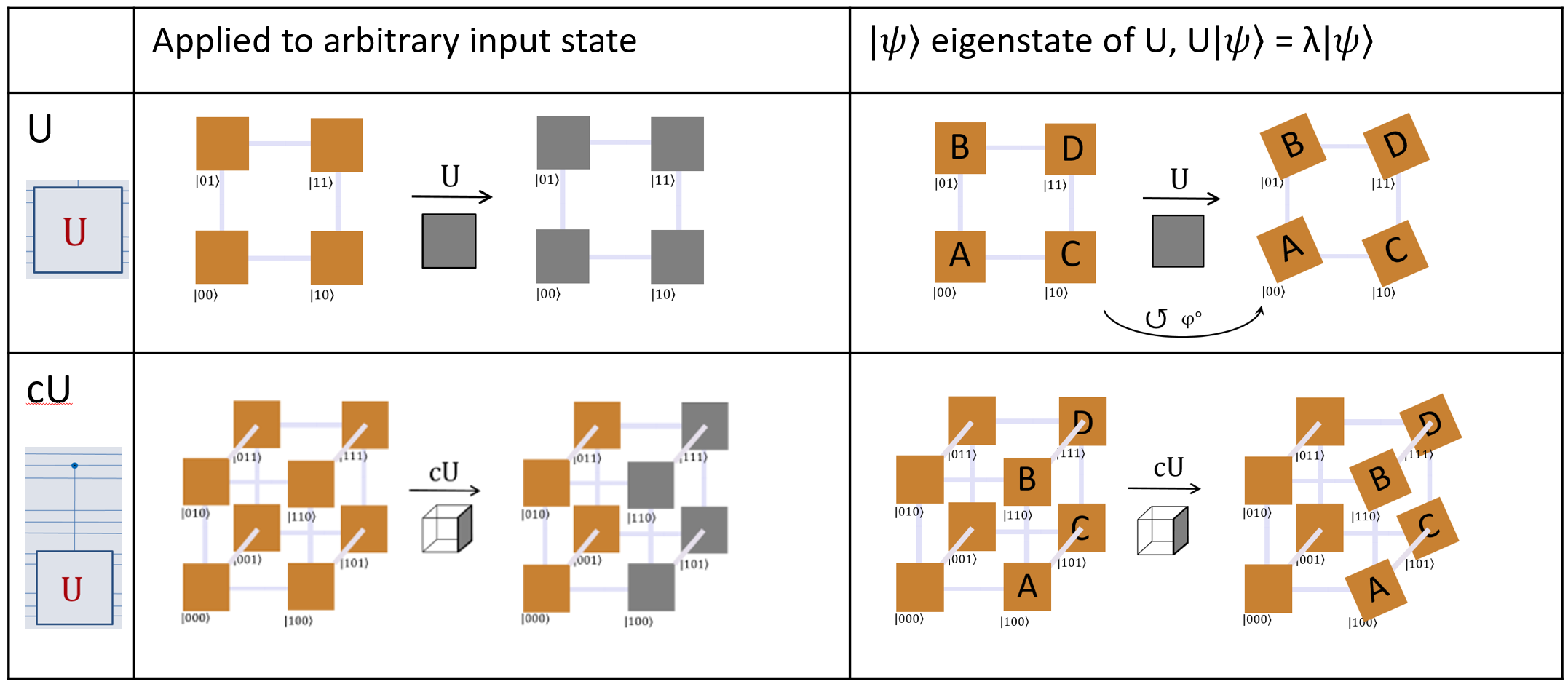}
\caption{Illustrating quantum phase kickback für a controlled-U. The unitary U operating on two qubits changes the amplitudes of a register from orange to grey (which could mean anything), but in the case of an eigenstate, this means turning all of them by the same angle. Controlled-U operating on three qubits applies U to the states where the control-qubit is 1, hence to those at the left-sided square of the cube. Applied to an eigenstate, controlled-U just turns the amplitudes of the left-sided square of the cube all by the same angle.}\label{CubeModel5}
\end{figure}

For the purpose of the quantum MOOCs at openHPI, the cube model for the
first time was applied to more advanced quantum algorithmic features, such
as quantum error correction, see Fig.~\ref{CubeModel3}, quantum oracles,
see Fig.~\ref{CubeModel4}, and quantum phase kickback, see Fig.~\ref{CubeModel5}.
 Based on the MOOCs, the visualization of some abstract algorithmic concepts with the quantum cube model is summarized and demonstrated in \cite{Justyoutube}.%

\end{appendices}

\bibliography{QuantumEducation_final}


\begin{thebibliography}{45}
\ifx \bisbn   \undefined \def \bisbn  #1{ISBN #1}\fi
\ifx \binits  \undefined \def \binits#1{#1}\fi
\ifx \bauthor  \undefined \def \bauthor#1{#1}\fi
\ifx \batitle  \undefined \def \batitle#1{#1}\fi
\ifx \bjtitle  \undefined \def \bjtitle#1{#1}\fi
\ifx \bvolume  \undefined \def \bvolume#1{\textbf{#1}}\fi
\ifx \byear  \undefined \def \byear#1{#1}\fi
\ifx \bissue  \undefined \def \bissue#1{#1}\fi
\ifx \bfpage  \undefined \def \bfpage#1{#1}\fi
\ifx \blpage  \undefined \def \blpage #1{#1}\fi
\ifx \burl  \undefined \def \burl#1{\textsf{#1}}\fi
\ifx \doiurl  \undefined \def \doiurl#1{\url{https://doi.org/#1}}\fi
\ifx \betal  \undefined \def \betal{\textit{et al.}}\fi
\ifx \binstitute  \undefined \def \binstitute#1{#1}\fi
\ifx \binstitutionaled  \undefined \def \binstitutionaled#1{#1}\fi
\ifx \bctitle  \undefined \def \bctitle#1{#1}\fi
\ifx \beditor  \undefined \def \beditor#1{#1}\fi
\ifx \bpublisher  \undefined \def \bpublisher#1{#1}\fi
\ifx \bbtitle  \undefined \def \bbtitle#1{#1}\fi
\ifx \bedition  \undefined \def \bedition#1{#1}\fi
\ifx \bseriesno  \undefined \def \bseriesno#1{#1}\fi
\ifx \blocation  \undefined \def \blocation#1{#1}\fi
\ifx \bsertitle  \undefined \def \bsertitle#1{#1}\fi
\ifx \bsnm \undefined \def \bsnm#1{#1}\fi
\ifx \bsuffix \undefined \def \bsuffix#1{#1}\fi
\ifx \bparticle \undefined \def \bparticle#1{#1}\fi
\ifx \barticle \undefined \def \barticle#1{#1}\fi
\bibcommenthead
\ifx \bconfdate \undefined \def \bconfdate #1{#1}\fi
\ifx \botherref \undefined \def \botherref #1{#1}\fi
\ifx \url \undefined \def \url#1{\textsf{#1}}\fi
\ifx \bchapter \undefined \def \bchapter#1{#1}\fi
\ifx \bbook \undefined \def \bbook#1{#1}\fi
\ifx \bcomment \undefined \def \bcomment#1{#1}\fi
\ifx \oauthor \undefined \def \oauthor#1{#1}\fi
\ifx \citeauthoryear \undefined \def \citeauthoryear#1{#1}\fi
\ifx \endbibitem  \undefined \def \endbibitem {}\fi
\ifx \bconflocation  \undefined \def \bconflocation#1{#1}\fi
\ifx \arxivurl  \undefined \def \arxivurl#1{\textsf{#1}}\fi
\csname PreBibitemsHook\endcsname

\bibitem[\protect\citeauthoryear{Greinert
  et~al.}{2023}]{franziska_greinert_towards_2023}
\begin{botherref}
\oauthor{\bsnm{Greinert}, \binits{F.}},
\oauthor{\bsnm{Müller}, \binits{R.}},
\oauthor{\bsnm{Goorney}, \binits{S.}},
\oauthor{\bsnm{Sherson}, \binits{J.}},
\oauthor{\bsnm{Ubben}, \binits{M.S.}}:
Towards a quantum ready workforce: the updated european competence framework
  for quantum technologies.
Frontiers in Quantum Science and Technology
\textbf{2}
(2023)
\doiurl{10.3389/frqst.2023.1225733}
\end{botherref}
\endbibitem

\bibitem[\protect\citeauthoryear{Bitzenbauer}{2021}]{philipp_bitzenbauer_quantum_2021}
\begin{botherref}
\oauthor{\bsnm{Bitzenbauer}, \binits{P.}}:
Quantum physics education research over the last two decades: A bibliometric
  analysis.
Education Sciences
\textbf{11}(11)
(2021)
\doiurl{10.3390/educsci11110699}
\end{botherref}
\endbibitem

\bibitem[\protect\citeauthoryear{Aiello
  et~al.}{2021}]{clarice_d_aiello_achieving_2021}
\begin{barticle}
\bauthor{\bsnm{Aiello}, \binits{C.D.}},
\bauthor{\bsnm{Awschalom}, \binits{D.D.}},
\bauthor{\bsnm{Bernien}, \binits{H.}},
\bauthor{\bsnm{Brower}, \binits{T.}},
\bauthor{\bsnm{Brown}, \binits{K.R.}},
\bauthor{\bsnm{Brun}, \binits{T.A.}},
\bauthor{\bsnm{Caram}, \binits{J.R.}},
\bauthor{\bsnm{Chitambar}, \binits{E.}},
\bauthor{\bsnm{Felice}, \binits{R.D.}},
\bauthor{\bsnm{Edmonds}, \binits{K.M.}},
\bauthor{\bsnm{Fox}, \binits{M.F.J.}},
\bauthor{\bsnm{Haas}, \binits{S.}},
\bauthor{\bsnm{Holleitner}, \binits{A.W.}},
\bauthor{\bsnm{Hudson}, \binits{E.R.}},
\bauthor{\bsnm{Hunt}, \binits{J.H.}},
\bauthor{\bsnm{Joynt}, \binits{R.}},
\bauthor{\bsnm{Koziol}, \binits{S.}},
\bauthor{\bsnm{Larsen}, \binits{M.}},
\bauthor{\bsnm{Lewandowski}, \binits{H.J.}},
\bauthor{\bsnm{McClure}, \binits{D.T.}},
\bauthor{\bsnm{Palsberg}, \binits{J.}},
\bauthor{\bsnm{Passante}, \binits{G.}},
\bauthor{\bsnm{Pudenz}, \binits{K.L.}},
\bauthor{\bsnm{Richardson}, \binits{C.J.K.}},
\bauthor{\bsnm{Rosenberg}, \binits{J.L.}},
\bauthor{\bsnm{Ross}, \binits{R.S.}},
\bauthor{\bsnm{Saffman}, \binits{M.}},
\bauthor{\bsnm{Singh}, \binits{M.}},
\bauthor{\bsnm{Steuerman}, \binits{D.W.}},
\bauthor{\bsnm{Stark}, \binits{C.}},
\bauthor{\bsnm{Thijssen}, \binits{J.}},
\bauthor{\bsnm{Vamivakas}, \binits{A.N.}},
\bauthor{\bsnm{Whitfield}, \binits{J.D.}},
\bauthor{\bsnm{Zwickl}, \binits{B.M.}}:
\batitle{Achieving a quantum smart workforce}.
\bjtitle{Quantum Science and Technology}
\bvolume{6}(\bissue{3}),
\bfpage{030501}
(\byear{2021})
\doiurl{10.1088/2058-9565/abfa64}
\end{barticle}
\endbibitem

\bibitem[\protect\citeauthoryear{Lee and
  Searles}{2021}]{kayla_lee_ibm-hbcu_2021}
\begin{bchapter}
\bauthor{\bsnm{Lee}, \binits{K.B.}},
\bauthor{\bsnm{Searles}, \binits{T.}}:
\bctitle{Ibm-hbcu quantum center: A model for industry-academic partnerships to
  advance the creation of a diverse, quantum aware workforce}.
In: \bbtitle{2021 IEEE International Conference on Quantum Computing and
  Engineering (QCE)},
pp. \bfpage{392}--\blpage{396}
(\byear{2021}).
\doiurl{10.1109/QCE52317.2021.00059}
\end{bchapter}
\endbibitem

\bibitem[\protect\citeauthoryear{Bungum and Selstø}{}]{bungum_what_2022}
\begin{botherref}
\oauthor{\bsnm{Bungum}, \binits{B.}},
\oauthor{\bsnm{Selstø}, \binits{S.}}:
What do quantum computing students need to know about quantum physics?
European Journal of Physics
\textbf{43}(5)
\doiurl{10.1088/1361-6404/ac7e8a}
\end{botherref}
\endbibitem

\bibitem[\protect\citeauthoryear{Stump
  et~al.}{2023}]{emily_m_stump_context_2023}
\begin{barticle}
\bauthor{\bsnm{Stump}, \binits{E.M.}},
\bauthor{\bsnm{Dew}, \binits{M.}},
\bauthor{\bsnm{Passante}, \binits{G.}},
\bauthor{\bsnm{Holmes}, \binits{N.G.}}:
\batitle{Context affects student thinking about sources of uncertainty in
  classical and quantum mechanics}.
\bjtitle{Phys. Rev. Phys. Educ. Res.}
\bvolume{19},
\bfpage{020157}
(\byear{2023})
\doiurl{10.1103/PhysRevPhysEducRes.19.020157}
\end{barticle}
\endbibitem

\bibitem[\protect\citeauthoryear{Meyer
  et~al.}{2022}]{josephine_c_meyer_todays_2022}
\begin{barticle}
\bauthor{\bsnm{Meyer}, \binits{J.C.}},
\bauthor{\bsnm{Passante}, \binits{G.}},
\bauthor{\bsnm{Pollock}, \binits{S.J.}},
\bauthor{\bsnm{Wilcox}, \binits{B.R.}}:
\batitle{Today's interdisciplinary quantum information classroom: Themes from a
  survey of quantum information science instructors}.
\bjtitle{Phys. Rev. Phys. Educ. Res.}
\bvolume{18},
\bfpage{010150}
(\byear{2022})
\doiurl{10.1103/PhysRevPhysEducRes.18.010150}
\end{barticle}
\endbibitem

\bibitem[\protect\citeauthoryear{Ahmed
  et~al.}{2022}]{shaeema_zaman_ahmed_student_2022}
\begin{barticle}
\bauthor{\bsnm{Ahmed}, \binits{S.Z.}},
\bauthor{\bsnm{Weidner}, \binits{C.A.}},
\bauthor{\bsnm{Jensen}, \binits{J.H.M.}},
\bauthor{\bsnm{Sherson}, \binits{J.F.}},
\bauthor{\bsnm{Lewandowski}, \binits{H.J.}}:
\batitle{Student use of a quantum simulation and visualization tool}.
\bjtitle{European Journal of Physics}
\bvolume{43}(\bissue{6}),
\bfpage{065703}
(\byear{2022})
\doiurl{10.1088/1361-6404/ac93c7}
\end{barticle}
\endbibitem

\bibitem[\protect\citeauthoryear{Delgado}{2023}]{francisco_delgado_extending_2023}
\begin{botherref}
\oauthor{\bsnm{Delgado}, \binits{F.}}:
Extending learning and collaboration in quantum information with internet
  support: A future perspective on research education beyond boundaries,
  limitations, and frontiers.
Future Internet
\textbf{15}(5)
(2023)
\doiurl{10.3390/fi15050160}
\end{botherref}
\endbibitem

\bibitem[\protect\citeauthoryear{Hasanovic}{2023}]{moamer_hasanovic_quantum_2023}
\begin{botherref}
\oauthor{\bsnm{Hasanovic}, \binits{M.}}:
Quantum education: How to teach a subject that nobody fully understands.
Seventeenth Conference on Education and Training in Optics and Photonics: ETOP
  2023,
1272331
(2023)
\doiurl{10.1117/12.2670468}
\end{botherref}
\endbibitem

\bibitem[\protect\citeauthoryear{Bondani
  et~al.}{2022}]{bondani_introducing_2022}
\begin{barticle}
\bauthor{\bsnm{Bondani}, \binits{M.}},
\bauthor{\bsnm{Chiofalo}, \binits{M.L.}},
\bauthor{\bsnm{Ercolessi}, \binits{E.}},
\bauthor{\bsnm{Macchiavello}, \binits{C.}},
\bauthor{\bsnm{Malgieri}, \binits{M.}},
\bauthor{\bsnm{Michelini}, \binits{M.}},
\bauthor{\bsnm{Mishina}, \binits{O.}},
\bauthor{\bsnm{Onorato}, \binits{P.}},
\bauthor{\bsnm{Pallotta}, \binits{F.}},
\bauthor{\bsnm{Satanassi}, \binits{S.}},
\bauthor{\bsnm{Stefanel}, \binits{A.}},
\bauthor{\bsnm{Sutrini}, \binits{C.}},
\bauthor{\bsnm{Testa}, \binits{I.}},
\bauthor{\bsnm{Zuccarini}, \binits{G.}}:
\batitle{Introducing quantum technologies at secondary school level: Challenges
  and potential impact of an online extracurricular course}.
\bjtitle{Physics}
\bvolume{4}(\bissue{4}),
\bfpage{1150}--\blpage{1167}
(\byear{2022})
\doiurl{10.3390/physics4040075}
\end{barticle}
\endbibitem

\bibitem[\protect\citeauthoryear{Freericks et~al.}{2019}]{Freericks_2017}
\begin{barticle}
\bauthor{\bsnm{Freericks}, \binits{J.K.}},
\bauthor{\bsnm{Cutler}, \binits{D.}},
\bauthor{\bsnm{Kruse}, \binits{A.}},
\bauthor{\bsnm{Vieira}, \binits{L.B.}}:
\batitle{{Teaching Quantum Mechanics to Over 28,000 Nonscientists}}.
\bjtitle{The Physics Teacher}
\bvolume{57}(\bissue{5}),
\bfpage{326}--\blpage{329}
(\byear{2019})
\doiurl{10.1119/1.5098924}
\end{barticle}
\endbibitem

\bibitem[\protect\citeauthoryear{Angara et~al.}{2020}]{Angara2020}
\begin{bchapter}
\bauthor{\bsnm{Angara}, \binits{P.P.}},
\bauthor{\bsnm{Stege}, \binits{U.}},
\bauthor{\bsnm{MacLean}, \binits{A.}}:
\bctitle{{Quantum Computing for High-School Students An Experience Report}}.
In: \bbtitle{{2020 IEEE International Conference on Quantum Computing and
  Engineering}}
(\byear{2020}).
\doiurl{10.1109/QCE49297.2020.00047}
\end{bchapter}
\endbibitem

\bibitem[\protect\citeauthoryear{Satanassi et~al.}{2022}]{Satanassi2022}
\begin{barticle}
\bauthor{\bsnm{Satanassi}, \binits{S.}},
\bauthor{\bsnm{Ercolessi}, \binits{E.}},
\bauthor{\bsnm{Levrini}, \binits{O.}}:
\batitle{Designing and implementing materials on quantum computing for
  secondary school students: The case of teleportation}.
\bjtitle{Phys. Rev. Phys. Educ. Res.}
\bvolume{18},
\bfpage{010122}
(\byear{2022})
\doiurl{10.1103/PhysRevPhysEducRes.18.010122}
\end{barticle}
\endbibitem

\bibitem[\protect\citeauthoryear{Satanassi}{2023}]{Satanassi2023}
\begin{botherref}
\oauthor{\bsnm{Satanassi}, \binits{S.}}:
{Investigating the learning potential of the Second Quantum Revolution:
  development of an approach for secondary school students}.
PhD thesis,
Bologna U.
(2023).
\doiurl{10.48676/unibo/amsdottorato/10716}
\end{botherref}
\endbibitem

\bibitem[\protect\citeauthoryear{Pospiech}{2021}]{Pospiech2021}
\begin{bbook}
\bauthor{\bsnm{Pospiech}, \binits{G.}}:
\bbtitle{Quantum Cryptography as an Approach for Teaching Quantum Physics},
pp. \bfpage{19}--\blpage{31}
(\byear{2021}).
\doiurl{10.1007/978-3-030-78720-2_2}
\end{bbook}
\endbibitem

\bibitem[\protect\citeauthoryear{Sutrini et~al.}{2022}]{Sutrini2022}
\begin{barticle}
\bauthor{\bsnm{Sutrini}, \binits{C.}},
\bauthor{\bsnm{Malgieri}, \binits{M.}},
\bauthor{\bsnm{Macchiavello}, \binits{C.}}:
\batitle{Quantum technologies: a course for teacher professional development.}
\bjtitle{Journal of Physics: Conference Series}
\bvolume{2297},
\bfpage{012018}
(\byear{2022})
\doiurl{10.1088/1742-6596/2297/1/012018}
\end{barticle}
\endbibitem

\bibitem[\protect\citeauthoryear{Sutrini et~al.}{2023}]{Sutrini2023a}
\begin{barticle}
\bauthor{\bsnm{Sutrini}, \binits{C.}},
\bauthor{\bsnm{Malgieri}, \binits{M.}},
\bauthor{\bsnm{Zuccarini}, \binits{G.}},
\bauthor{\bsnm{Macchiavello}, \binits{C.}}:
\batitle{A teacher professional development course on quantum technologies:
  discussion of results}.
\bjtitle{Journal of Physics: Conference Series}
\bvolume{2490},
\bfpage{012006}
(\byear{2023})
\doiurl{10.1088/1742-6596/2490/1/012006}
\end{barticle}
\endbibitem

\bibitem[\protect\citeauthoryear{Araceli}{}]{araceli_venegasgomez_quantum_2020}
\begin{botherref}
\oauthor{\bsnm{Araceli}, \binits{V.-G.}}:
The quantum ecosystem and its future workforce.
Photonic Views
\textbf{17}(6),
34--38
\doiurl{10.1002/phvs.202000044}
\end{botherref}
\endbibitem

\bibitem[\protect\citeauthoryear{Plunkett
  et~al.}{2020}]{thomas_plunkett_survey_2020}
\begin{bchapter}
\bauthor{\bsnm{Plunkett}, \binits{T.}},
\bauthor{\bsnm{Frantz}, \binits{T.L.}},
\bauthor{\bsnm{Khatri}, \binits{H.}},
\bauthor{\bsnm{Rajendran}, \binits{P.}},
\bauthor{\bsnm{Midha}, \binits{S.}}:
\bctitle{A survey of educational efforts to accelerate a growing quantum
  workforce}.
In: \bbtitle{2020 IEEE International Conference on Quantum Computing and
  Engineering (QCE)},
pp. \bfpage{330}--\blpage{336}
(\byear{2020}).
\doiurl{10.1109/QCE49297.2020.00048}
\end{bchapter}
\endbibitem

\bibitem[\protect\citeauthoryear{Fox
  et~al.}{2020}]{michael_f_j_fox_preparing_2020}
\begin{barticle}
\bauthor{\bsnm{Fox}, \binits{M.F.J.}},
\bauthor{\bsnm{Zwickl}, \binits{B.M.}},
\bauthor{\bsnm{Lewandowski}, \binits{H.J.}}:
\batitle{Preparing for the quantum revolution: What is the role of higher
  education?}
\bjtitle{Phys. Rev. Phys. Educ. Res.}
\bvolume{16},
\bfpage{020131}
(\byear{2020})
\doiurl{10.1103/PhysRevPhysEducRes.16.020131}
\end{barticle}
\endbibitem

\bibitem[\protect\citeauthoryear{Asfaw
  et~al.}{2022}]{abraham_asfaw_building_2022}
\begin{barticle}
\bauthor{\bsnm{Asfaw}, \binits{A.}},
\bauthor{\bsnm{Blais}, \binits{A.}},
\bauthor{\bsnm{Brown}, \binits{K.R.}},
\bauthor{\bsnm{Candelaria}, \binits{J.}},
\bauthor{\bsnm{Cantwell}, \binits{C.}},
\bauthor{\bsnm{Carr}, \binits{L.D.}},
\bauthor{\bsnm{Combes}, \binits{J.}},
\bauthor{\bsnm{Debroy}, \binits{D.M.}},
\bauthor{\bsnm{Donohue}, \binits{J.M.}},
\bauthor{\bsnm{Economou}, \binits{S.E.}},
\bauthor{\bsnm{Edwards}, \binits{E.}},
\bauthor{\bsnm{Fox}, \binits{M.F.J.}},
\bauthor{\bsnm{Girvin}, \binits{S.M.}},
\bauthor{\bsnm{Ho}, \binits{A.}},
\bauthor{\bsnm{Hurst}, \binits{H.M.}},
\bauthor{\bsnm{Jacob}, \binits{Z.}},
\bauthor{\bsnm{Johnson}, \binits{B.R.}},
\bauthor{\bsnm{Johnston-Halperin}, \binits{E.}},
\bauthor{\bsnm{Joynt}, \binits{R.}},
\bauthor{\bsnm{Kapit}, \binits{E.}},
\bauthor{\bsnm{Klein-Seetharaman}, \binits{J.}},
\bauthor{\bsnm{Laforest}, \binits{M.}},
\bauthor{\bsnm{Lewandowski}, \binits{H.J.}},
\bauthor{\bsnm{Lynn}, \binits{T.W.}},
\bauthor{\bsnm{McRae}, \binits{C.R.H.}},
\bauthor{\bsnm{Merzbacher}, \binits{C.}},
\bauthor{\bsnm{Michalakis}, \binits{S.}},
\bauthor{\bsnm{Narang}, \binits{P.}},
\bauthor{\bsnm{Oliver}, \binits{W.D.}},
\bauthor{\bsnm{Palsberg}, \binits{J.}},
\bauthor{\bsnm{Pappas}, \binits{D.P.}},
\bauthor{\bsnm{Raymer}, \binits{M.G.}},
\bauthor{\bsnm{Reilly}, \binits{D.J.}},
\bauthor{\bsnm{Saffman}, \binits{M.}},
\bauthor{\bsnm{Searles}, \binits{T.A.}},
\bauthor{\bsnm{Shapiro}, \binits{J.H.}},
\bauthor{\bsnm{Singh}, \binits{C.}}:
\batitle{Building a quantum engineering undergraduate program}.
\bjtitle{IEEE Transactions on Education}
\bvolume{65}(\bissue{2}),
\bfpage{220}--\blpage{242}
(\byear{2022})
\doiurl{10.1109/TE.2022.3144943}
\end{barticle}
\endbibitem

\bibitem[\protect\citeauthoryear{Gerke
  et~al.}{2022}]{f_gerke_requirements_2022}
\begin{barticle}
\bauthor{\bsnm{Gerke}, \binits{F.}},
\bauthor{\bsnm{Müller}, \binits{R.}},
\bauthor{\bsnm{Bitzenbauer}, \binits{P.}},
\bauthor{\bsnm{Ubben}, \binits{M.}},
\bauthor{\bsnm{Weber}, \binits{K.-A.}}:
\batitle{Requirements for future quantum workforce – a delphi study}.
\bjtitle{Journal of Physics: Conference Series}
\bvolume{2297}(\bissue{1}),
\bfpage{012017}
(\byear{2022})
\doiurl{10.1088/1742-6596/2297/1/012017}
\end{barticle}
\endbibitem

\bibitem[\protect\citeauthoryear{Greinert
  et~al.}{2023}]{franziska_greinert_future_2023}
\begin{barticle}
\bauthor{\bsnm{Greinert}, \binits{F.}},
\bauthor{\bsnm{M\"uller}, \binits{R.}},
\bauthor{\bsnm{Bitzenbauer}, \binits{P.}},
\bauthor{\bsnm{Ubben}, \binits{M.S.}},
\bauthor{\bsnm{Weber}, \binits{K.-A.}}:
\batitle{Future quantum workforce: Competences, requirements, and forecasts}.
\bjtitle{Phys. Rev. Phys. Educ. Res.}
\bvolume{19},
\bfpage{010137}
(\byear{2023})
\doiurl{10.1103/PhysRevPhysEducRes.19.010137}
\end{barticle}
\endbibitem

\bibitem[\protect\citeauthoryear{Combarro
  et~al.}{}]{elias_f_combarro_report_2021}
\begin{botherref}
\oauthor{\bsnm{Combarro}, \binits{E.F.}},
\oauthor{\bsnm{Vallecorsa}, \binits{S.}},
\oauthor{\bsnm{Rodríguez-Muñiz}, \binits{L.J.}},
\oauthor{\bsnm{Aguilar-González}, \binits{A.}},
\oauthor{\bsnm{Ranilla}, \binits{J.R.}},
\oauthor{\bsnm{Di~Meglio}, \binits{A.}}:
A report on teaching a series of online lectures on quantum computing from
  {CERN}.
The Journal of Supercomputing
\textbf{77},
14405--14435
\doiurl{10.1007/s11227-021-03847-9}
\end{botherref}
\endbibitem

\bibitem[\protect\citeauthoryear{Maldonado-Romo and
  Yeh}{2022}]{alberto_maldonado-romo_quantum_2022}
\begin{bchapter}
\bauthor{\bsnm{Maldonado-Romo}, \binits{A.}},
\bauthor{\bsnm{Yeh}, \binits{L.}}:
\bctitle{Quantum computing online workshops and hackathon for spanish speakers:
  A case study}.
In: \bbtitle{2022 IEEE International Conference on Quantum Computing and
  Engineering (QCE)},
pp. \bfpage{709}--\blpage{717}.
\bpublisher{IEEE Computer Society},
\blocation{Los Alamitos, CA, USA}
(\byear{2022}).
\doiurl{10.1109/QCE53715.2022.00096}
\end{bchapter}
\endbibitem

\bibitem[\protect\citeauthoryear{Aithal}{2023}]{p_s_aithal_advances_2023}
\begin{barticle}
\bauthor{\bsnm{Aithal}, \binits{P.S.}}:
\batitle{Advances and new research opportunities in quantum computing
  technology by integrating it with other icct underlying technologies}.
\bjtitle{International Journal of Case Studies in Business, IT and Education
  (IJCSBE)}
\bvolume{7}(\bissue{3}),
\bfpage{314}--\blpage{358}
(\byear{2023})
\doiurl{10.47992/IJCSBE.2581.6942.0304}
\end{barticle}
\endbibitem

\bibitem[\protect\citeauthoryear{Serth et~al.}{2022}]{Serth_2022}
\begin{botherref}
\oauthor{\bsnm{Serth}, \binits{S.}},
\oauthor{\bsnm{Staubitz}, \binits{T.}},
\oauthor{\bsnm{Elten}, \binits{M.}},
\oauthor{\bsnm{Meinel}, \binits{C.}}:
Measuring the effects of course modularizations in online courses for life-long
  learners.
Frontiers in Education
\textbf{7}
(2022)
\doiurl{10.3389/feduc.2022.1008545}
\end{botherref}
\endbibitem

\bibitem[\protect\citeauthoryear{Just}{2020}]{Just2020}
\begin{bbook}
\bauthor{\bsnm{Just}, \binits{B.}}:
\bbtitle{Quantencomputing {k}ompakt: Spukhafte Fernwirkung und Teleportation
  {e}ndlich {v}erständlich}.
\bpublisher{Springer},
\blocation{Berlin, Heidelberg}
(\byear{2020}).
\doiurl{10.1007/978-3-662-61889-9}
\end{bbook}
\endbibitem

\bibitem[\protect\citeauthoryear{Just}{2022}]{Just2022}
\begin{bbook}
\bauthor{\bsnm{Just}, \binits{B.}}:
\bbtitle{Quantum Computing Compact: Spooky Action at a Distance and
  Teleportation Easy to Understand}.
\bpublisher{Springer},
\blocation{Berlin, Heidelberg}
(\byear{2022}).
\doiurl{10.1007/978-3-662-65008-0}
\end{bbook}
\endbibitem

\bibitem[\protect\citeauthoryear{Seegerer et~al.}{2021}]{seegerer2021quantum}
\begin{bchapter}
\bauthor{\bsnm{Seegerer}, \binits{S.}},
\bauthor{\bsnm{Michaeli}, \binits{T.}},
\bauthor{\bsnm{Romeike}, \binits{R.}}:
\bctitle{Quantum computing as a topic in computer science education}.
In: \bbtitle{Proceedings of the 16th Workshop in Primary and Secondary
  Computing Education}.
\bsertitle{WiPSCE '21}.
\bpublisher{Association for Computing Machinery},
\blocation{New York, NY, USA}
(\byear{2021}).
\doiurl{10.1145/3481312.3481348}
\end{bchapter}
\endbibitem

\bibitem[\protect\citeauthoryear{Mueller-Roemer et~al.}{2022}]{Fraunhofer2022}
\begin{botherref}
\oauthor{\bsnm{Mueller-Roemer}, \binits{J.S.}},
\oauthor{\bsnm{28Smiles}},
\oauthor{\bsnm{Wagner}, \binits{P.}}:
QCVIS — Interactive Visualization of Quantum Algorithms.
Availabe at \url{https://github.com/fh-igd-iet/qcvis}
(2022)
\end{botherref}
\endbibitem

\bibitem[\protect\citeauthoryear{Bley et~al.}{2024}]{bley2024visualizing}
\begin{botherref}
\oauthor{\bsnm{Bley}, \binits{J.}},
\oauthor{\bsnm{Rexigel}, \binits{E.}},
\oauthor{\bsnm{Arias}, \binits{A.}},
\oauthor{\bsnm{Longen}, \binits{N.}},
\oauthor{\bsnm{Krupp}, \binits{L.}},
\oauthor{\bsnm{Kiefer-Emmanouilidis}, \binits{M.}},
\oauthor{\bsnm{Lukowicz}, \binits{P.}},
\oauthor{\bsnm{Donhauser}, \binits{A.}},
\oauthor{\bsnm{Küchemann}, \binits{S.}},
\oauthor{\bsnm{Kuhn}, \binits{J.}},
\oauthor{\bsnm{Widera}, \binits{A.}}:
Visualizing Entanglement in multi-Qubit Systems.
Preprint at \url{https://arxiv.org/abs/2305.07596v4}
(2024)
\end{botherref}
\endbibitem

\bibitem[\protect\citeauthoryear{{Qiskit contributors}}{2023}]{Qiskit}
\begin{botherref}
\oauthor{\bsnm{{Qiskit contributors}}}:
Qiskit: An Open-source Framework for Quantum Computing
(2023).
\doiurl{10.5281/zenodo.2573505}
\end{botherref}
\endbibitem

\bibitem[\protect\citeauthoryear{Plattner et~al.}{2010}]{DesignThinking}
\begin{bbook}
\bauthor{\bsnm{Plattner}, \binits{H.}},
\bauthor{\bsnm{Meinel}, \binits{C.}},
\bauthor{\bsnm{Leifer}, \binits{L.}}:
\bbtitle{Design Thinking: Understand - Improve - Apply}.
\bpublisher{Springer},
\blocation{Heidelberg}
(\byear{2010}).
\doiurl{10.1007/978-3-642-13757-0}
\end{bbook}
\endbibitem

\bibitem[\protect\citeauthoryear{Bennett and Brassard}{1984}]{Bennet}
\begin{bchapter}
\bauthor{\bsnm{Bennett}, \binits{C.}},
\bauthor{\bsnm{Brassard}, \binits{G.}}:
\bctitle{Withdrawn: Quantum cryptography: Public key distribution and coin
  tossing},
vol. \bseriesno{560},
pp. \bfpage{175}--\blpage{179}
(\byear{1984}).
\doiurl{10.1016/j.tcs.2011.08.039}
\end{bchapter}
\endbibitem

\bibitem[\protect\citeauthoryear{Clauser et~al.}{1969}]{CHSH}
\begin{barticle}
\bauthor{\bsnm{Clauser}, \binits{J.F.}},
\bauthor{\bsnm{Horne}, \binits{M.A.}},
\bauthor{\bsnm{Shimony}, \binits{A.}},
\bauthor{\bsnm{Holt}, \binits{R.A.}}:
\batitle{Proposed experiment to test local hidden-variable theories}.
\bjtitle{Phys. Rev. Lett.}
\bvolume{23},
\bfpage{880}--\blpage{884}
(\byear{1969})
\doiurl{10.1103/PhysRevLett.23.880}
\end{barticle}
\endbibitem

\bibitem[\protect\citeauthoryear{Cleve et~al.}{2004}]{CHSH_Game}
\begin{bchapter}
\bauthor{\bsnm{Cleve}, \binits{R.}},
\bauthor{\bsnm{Hoyer}, \binits{P.}},
\bauthor{\bsnm{Toner}, \binits{B.}},
\bauthor{\bsnm{Watrous}, \binits{J.}}:
\bctitle{Consequences and limits of nonlocal strategies}.
In: \bbtitle{Proceedings. 19th IEEE Annual Conference on Computational
  Complexity, 2004.},
pp. \bfpage{236}--\blpage{249}
(\byear{2004}).
\doiurl{10.1109/CCC.2004.1313847}
\end{bchapter}
\endbibitem

\bibitem[\protect\citeauthoryear{Nielsen and Chuang}{2011}]{NielsenChuang}
\begin{bbook}
\bauthor{\bsnm{Nielsen}, \binits{M.A.}},
\bauthor{\bsnm{Chuang}, \binits{I.L.}}:
\bbtitle{Quantum Computation and Quantum Information: 10th Anniversary
  Edition}.
\bpublisher{Cambridge University Press},
\blocation{Cambridge}
(\byear{2011}).
\doiurl{10.1017/CBO9780511976667}
\end{bbook}
\endbibitem

\bibitem[\protect\citeauthoryear{Shor}{1997}]{Shor}
\begin{barticle}
\bauthor{\bsnm{Shor}, \binits{P.W.}}:
\batitle{Polynomial-time algorithms for prime factorization and discrete
  logarithms on a quantum computer}.
\bjtitle{SIAM Journal on Computing}
\bvolume{26}(\bissue{5}),
\bfpage{1484}--\blpage{1509}
(\byear{1997})
\doiurl{10.1137/S0097539795293172}
\end{barticle}
\endbibitem

\bibitem[\protect\citeauthoryear{Harrow et~al.}{2009}]{HHL}
\begin{barticle}
\bauthor{\bsnm{Harrow}, \binits{A.W.}},
\bauthor{\bsnm{Hassidim}, \binits{A.}},
\bauthor{\bsnm{Lloyd}, \binits{S.}}:
\batitle{Quantum algorithm for linear systems of equations}.
\bjtitle{Phys. Rev. Lett.}
\bvolume{103},
\bfpage{150502}
(\byear{2009})
\doiurl{10.1103/PhysRevLett.103.150502}
\end{barticle}
\endbibitem

\bibitem[\protect\citeauthoryear{Preskill}{2018}]{Preskill2018}
\begin{barticle}
\bauthor{\bsnm{Preskill}, \binits{J.}}:
\batitle{Quantum {C}omputing in the {NISQ} era and beyond}.
\bjtitle{{Quantum}}
\bvolume{2},
\bfpage{79}
(\byear{2018})
\doiurl{10.22331/q-2018-08-06-79}
\end{barticle}
\endbibitem

\bibitem[\protect\citeauthoryear{Jacquier et~al.}{2022}]{jacquier2022}
\begin{bbook}
\bauthor{\bsnm{Jacquier}, \binits{A.}},
\bauthor{\bsnm{Kondratyev}, \binits{O.}},
\bauthor{\bsnm{Lipton}, \binits{A.}},
\bauthor{\bsnm{Prado}, \binits{M.L.}}:
\bbtitle{Quantum Machine Learning and Optimisation in Finance: On the Road to
  Quantum Advantage}.
\bpublisher{Packt Publishing Ltd.},
\blocation{Birmingham, UK}
(\byear{2022}).
\burl{https://www.packtpub.com/product/quantum-machine-learning-and-optimisation-in-finance/9781801813570}
\end{bbook}
\endbibitem

\bibitem[\protect\citeauthoryear{Just}{2023}]{Justyoutube}
\begin{botherref}
\oauthor{\bsnm{Just}, \binits{B.}}:
The Algorithmic Principle of Quantum Phase Kickback (2023).
Youtube. Availabe at \url{https://www.youtube.com/watch?v=OzDJpeUhwvc}
\end{botherref}
\endbibitem

\bibitem[\protect\citeauthoryear{Feynman}{1985}]{Feynman}
\begin{bbook}
\bauthor{\bsnm{Feynman}, \binits{R.P.}}:
\bbtitle{QED, The Strange Theory of Light and Matter}.
\bpublisher{Princeton University Press},
\blocation{Princeton}
(\byear{1985}).
\doiurl{10.2307/j.ctt2jc8td.}
\end{bbook}
\endbibitem

\end{thebibliography}
\end{document}